\documentclass[aps, prl, twocolumn, superscriptaddress]{revtex4}

\usepackage{graphicx}
\usepackage[utf8]{inputenc}
\usepackage[english]{babel}
\usepackage{amsmath}
\usepackage{amssymb}
\usepackage{xcolor}
\usepackage[colorlinks=true,
linkcolor=blue,
citecolor=red,
urlcolor=blue,
hypertexnames=true]
{hyperref}
\usepackage{url}

\newcommand{\blue}{\textcolor{black}}

\newcommand{\bea}{\begin{eqnarray}}
\newcommand{\eea}{\end{eqnarray}}
\newcommand{\be}{\begin{equation}}
\newcommand{\ee}{\end{equation}}

\begin{document}

\title{Emergence of Correlations Between Records and Maximum in Random Time Series}

\author{Satya N. Majumdar}
\affiliation{Universit\'e Paris-Saclay, CNRS, LPTMS, 91405 Orsay, France}

\author{Sanjib Sabhapandit}
\affiliation{Raman Research Institute, Bengaluru 560080, India}

\author{Gr\'egory Schehr}
\affiliation{Laboratoire de Physique Th\'eorique et Hautes \'Energies, Sorbonne Universit\'e and CNRS, 4 place Jussieu, 75252 Paris Cedex 05, France}

\date{\today}

\begin{abstract}
We compute exactly the covariance between the maximum $M$ and the number of
records $K$ in a discrete time series of $n$ entries as a quantitative probe of the history leading to an extreme event.
For \emph{continuous} IID series, with a common probability distribution function (PDF) $p(x)$, we \blue{find} the striking result 
that $M$ and $K$ are uncorrelated for any $n \geq 1$. We show that this independence is rather fragile and the covariance $C_q(n)$
between $M$ and $K$ becomes nonzero when $p(x)$ becomes discontinuous due to a delta-component at $x=0$ with an amplitude $0< q < 1$. This time series models the occurrence of dry days with probability $q$ in a
precipitation time series when there is no rainfall. 
We show that there is an interesting shape transition at a critical value $q_c$ where the covariance $C_q(n)$, as a function of
$n$, changes from monotonic ($q<q_c$) to non monotonic behavior (for $q>q_c$). We \blue{compute} the critical strength $q_c$
exactly. Furthermore, for large $n$, we \blue{show} that $C_q(n)$ decays as a power law, with an exponent that depends on the tail
of $p(x)$. We then \blue{compare} our theoretical predictions with real precipitation data from several weather stations across Germany, finding
a very good agreement. 
\end{abstract}

\maketitle

\emph{Introduction.}
Extreme events arise in a wide variety of contexts, including climate science,
hydrology, finance, ecology and sports~\cite{Gumbel58,Lead82,arnold,Resnick08,Review_EVS,Book_OUP,Fis25}. In all these situations, one typically has
a time series of fixed duration. This time series may represent the daily precipitation or the daily temperature recorded in a given weather station. The statistics of extremes in such a time series play a crucial role in characterizing the global warming and climate changes~\cite{SK99,Fis25,Easter,Redner06,WK10,Hansen2012,rainfall,drought}. One of the 
simplest observables characterising the extreme fluctuations in these contexts is the value of the 
global maximum reached over a fixed time interval. The other important observable is the number of records
in that interval: a record occurs at a given
time if the value of the time series at that time exceeds all previous
values~\cite{arnold,record-review,GMS17,Book_OUP}. 
As more records occur, the global maximum increases: in fact, the global maximum is simply the value of the last record in that interval. \blue{Yet, neither the value of the global maximum nor the number of records reveal how this} global maximum was reached: did it emerge through many successive
record-breaking events, or from a single exceptional jump after a long quiet period? The relationship between the number of records
and the value of the global maximum clearly depends on the history of the time series. For example, if the global maximum has a large value,
one may naively expect that the number of records needed to be broken to reach this maximum is also large. Similarly, a lower value
of the maximum may indicate a small number of records. This suggests, naively, that these two observables are positively correlated.
Their correlation, of course, is encoded in their joint statistics. While maxima and records have each been extensively studied separately 
over the past decades (for a recent review, see \cite{Book_OUP}), their joint statistics have received remarkably little attention.

\begin{figure}[t]
\centering
\includegraphics[width=\linewidth]{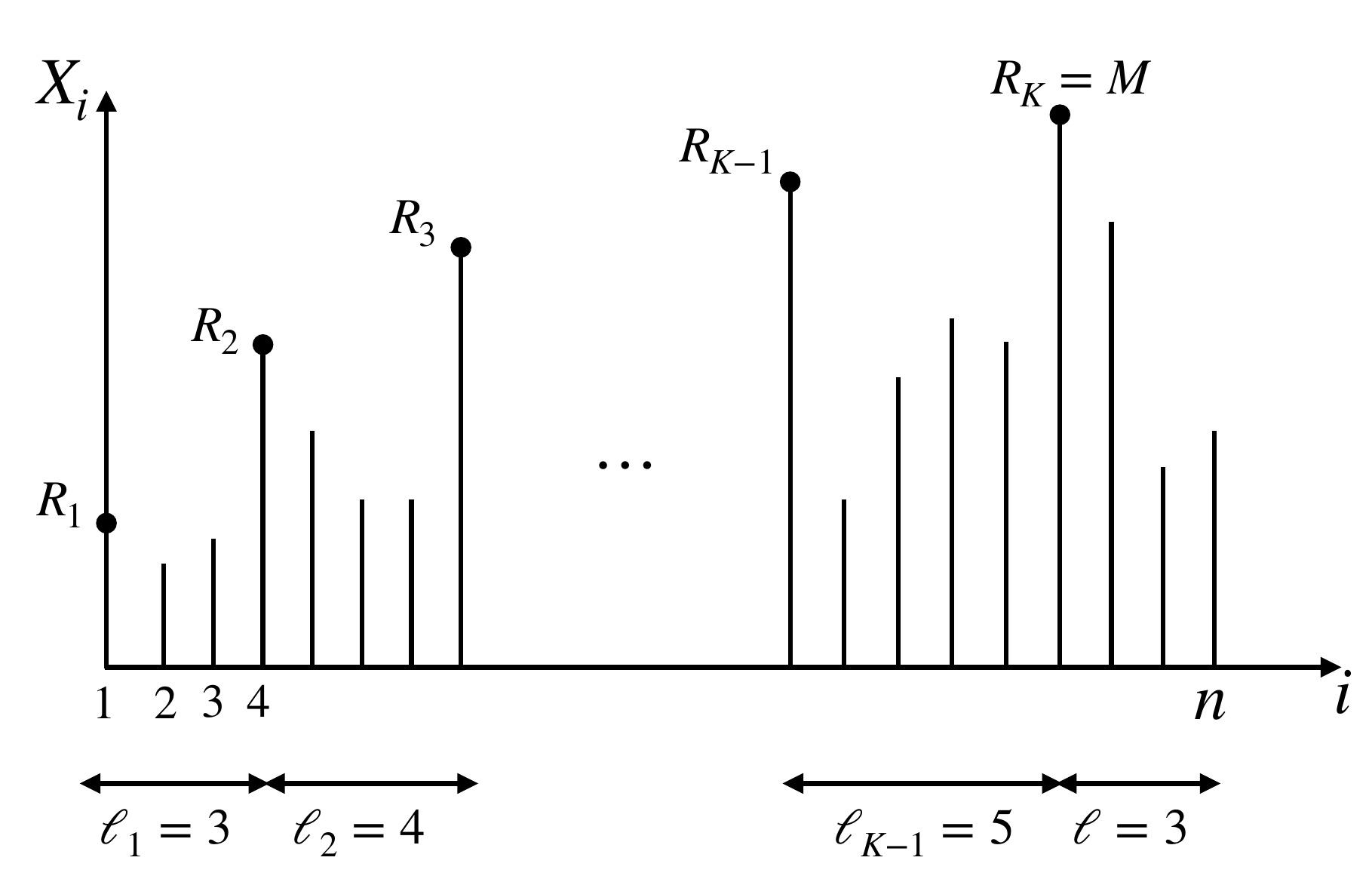}
\caption{A typical sequence of IID discrete time series with $n$ entries and $K$ records. Here $R_i$'s denote the record values, with $R_K = M$ denoting the global maximum, which is also the last record value. The $\ell_i$'s denote the record ages, i.e., the interval between successive records. The interval following the last record is denoted by $\ell$.}
\label{fig:schematic}
\end{figure}

In this Letter, we first study the joint statistics of the global maximum and the number of records in the simplest time series, namely 
a discrete sequence of $n$ entries $X_i$'s that are independent and identically distributed (IID) via a common {probability} distribution 
function (PDF) $p(x)$ -- see Fig. \ref{fig:schematic}. We will first consider the case where $p(x)$ is continuous. For such an IID time series, let $M = {\max} \{X_1, X_2,\cdots, X_n\}$ denote the global maximum value up to $n$ steps and $K$ denote the number of records, both being random variables, fluctuating from sample to sample. 
In this IID model, the statistics of both $M$ and $K$, individually, have been studied extensively, with many applications~\cite{Gumbel58,Lead82,arnold,Resnick08,Review_EVS,Book_OUP,SK99,Fis25,Easter,Redner06,WK10,Hansen2012,rainfall,drought,record-review,GMS17}. Interestingly, both these observables exhibit universal scaling behaviors that are independent of the details of $p(x)$. For instance, the maximum $M$, suitably centered and scaled, converges, in the large $n$ limit, to one of the three canonical extreme value distributions known as  Fr{\'e}chet, Gumbel, and Weibull, depending only on the tail of $p(x)$ for large $x$~\cite{Gumbel58,Lead82,arnold,Resnick08,Review_EVS,Book_OUP}. This universality, i.e., the lack of dependence on \blue{the details of} $p(x)$, is even stronger for the number of records $K$ and holds for any $n$, and not just for large $n$. For example, the average number of records is given by $\langle K\rangle = \sum_{k=1}^n 1/k$ for any $n$ and for arbitrary continuous $p(x)$. 
For large $n$ it increases very slowly as  $\langle K\rangle\sim\ln n$, indicating that records become harder and harder to break (see e.g.~\cite{Review_EVS,Book_OUP}). This naturally raises the question: are $M$ and $K$ positively correlated, as naively expected, for this IID sequence with a continuous $p(x)$? Our first result in this Letter is rather striking and unexpected: we find that $M$ and $K$ are completely uncorrelated for any $n$ and for any continuous $p(x)$. This means that one cannot infer any information on the statistics of $M$ by knowing that of $K$, or vice versa.

\begin{figure}[t]
\centering
\includegraphics[width=0.49\linewidth]{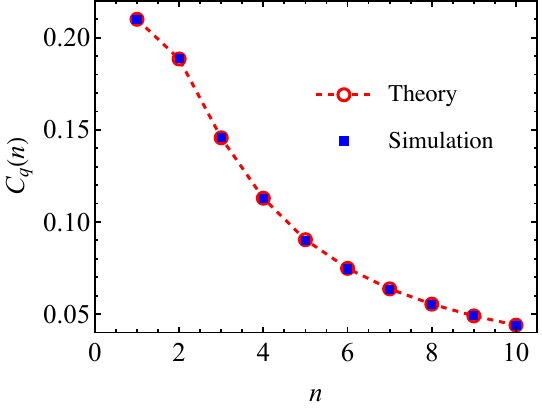}
\includegraphics[width=0.49\linewidth]{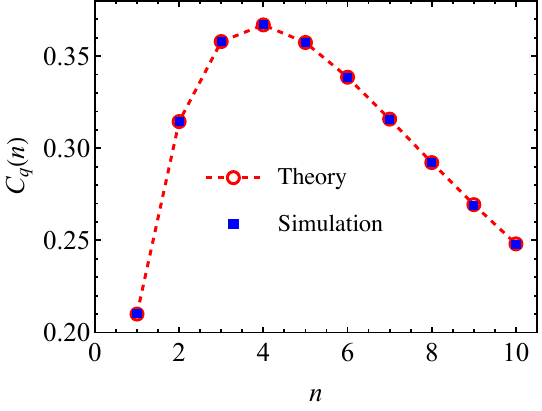}
\caption{Simulations showing $C_q(n)$ vs $n$ for $q=0.3$ (left panel) and $q=0.7$ (right panel) for the exponential distribution $f(x) = e^{-x}$ for $x>0$. For $q<q_c = 0.372281 \ldots$, the covariance $C_q(n)$ decays monotonically with increasing $n$, while for $q>q_c$, it shows a nonmonotonic behavior. The points show the simulation results while the dashed line corresponds to exact analytical results obtained from the small $v$ expansion of Eq. (\ref{GF_Cq_final}).}\label{Fig_Exp}
\end{figure}

\begin{figure}[t]
\centering
\includegraphics[width=0.49\linewidth]{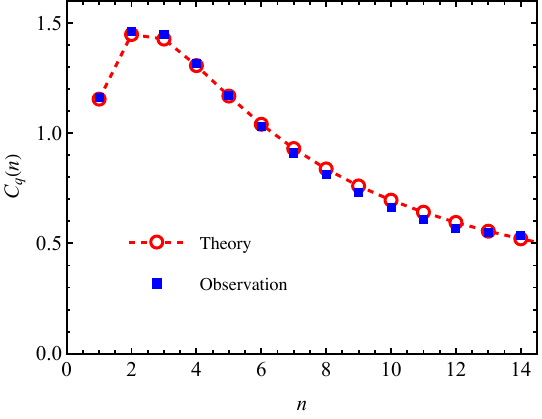}
\includegraphics[width=0.49\linewidth]{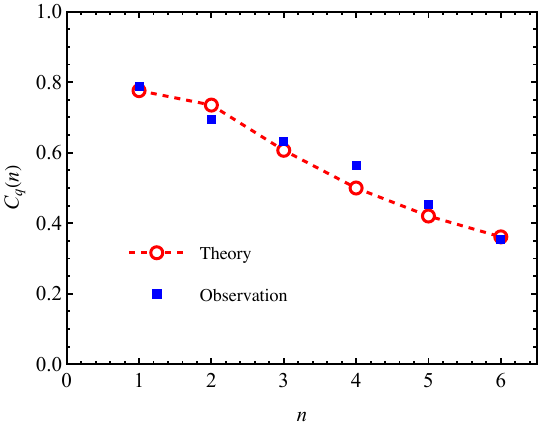}
\caption{Daily precipitation data from DWD compared with the IID dry-day model.
Left: stations with dry-day fraction close to $q=0.501569\ldots$, above the
empirical threshold $q_c=0.3152\ldots$, show a nonmonotonic covariance.  Right:
November data with $q=0.232353\ldots<q_c=0.279499\ldots$ show a monotone
decrease.  The theoretical curves use the empirical wet-day distribution.}
\label{fig:dwd}
\end{figure}

This then naturally leads to another interesting question: how robust is this super-universality of this lack of correlation between $K$ and $M$? Is it purely accidental and is valid only for IID sequences with a continuous $p(x)$? To investigate this question, one can either break the continuity assumption on $p(x)$, or introduce correlations between the entries. Both scenarios appear quite naturally in many practical situations. 
In this Letter, our primary focus will be on the first scenario and we will discuss the latter scenario briefly in the conclusion. 
The first scenario, where the entries are IID but the distribution $p(x)$ is non-continuous, arises frequently in analysing the rain-fall time-series. In this case the entry $X_i \geq 0$ represents the amount of precipitation on day $i$. However, there are ``dry'' days when there is no rain-fall, indicating that $X_i = 0$ for a dry-day (see Fig.  \ref{Fig_intro_dry}). Thus, in the distribution $p(x)$ of the individual entry, there is a finite probability $q$ that $X_i = 0$, where $0< q < 1$ represents the fraction of the dry days. This means that, effectively, the distribution $p(x)$, supported over $x \in [0, \infty)$, now acquires a delta-function at $x=0$ with weight $q$, in addition to having a smooth continuous part with weight $(1-q)$,~i.e., 
\begin{equation}
p(x)=q\,\delta(x)+(1-q)\,f(x) \;,
\label{eq:mixed}
\end{equation}
where $f(x)$ represents a smooth continuous PDF, with the normalisation $\int_{0}^\infty f(x)\, dx = 1$. This is known as the Bernoulli model of time-series with dry-days~\cite{Westcott77, BN92, DGP08, PK08,MBK19}. The presence of nonzero $q>0$ does not affect the statistics of the maximum $M$ for large $n$. Similarly the \blue{logarithmic growth} of the average number of records \blue{with $n$} remains unaffected by a nonzero $q$, to leading order for large $n$ \cite{MBK19}. However, in this Letter, we show that the presence of nonzero $q>0$ has a rich and profound effect on the correlations between $M$ and $K$. This is reflected in a nontrivial behavior of the covariance 
\begin{equation}
C_q(n) =\langle M K\rangle-\langle M\rangle\langle K\rangle \;.
\label{eq:defCN}
\end{equation}
We show that its dependence on $n$ undergoes a qualitative change at a critical dry-day
probability $q_c$. For $q<q_c$, the covariance $C_q(n)$ decreases monotonically with increasing
$n$, whereas for $q>q_c$ it first increases, reaches a maximum, and then
decreases [see Fig.~\ref{Fig_Exp} for the case of an exponential distribution $f(x) = e^{-x}$ for $x>0$]. We compute $q_c$ exactly for arbitrary $f(x)$. The physical mechanism behind this
``shape transition'' at $q=q_c$ is very interesting and can be qualitatively understood as follows. When $q$ is small, there are only very few dry days and the time series for large $n$ consists of long patches of only wet days. Hence, as $n$ increases, the time series resembles more and more the one without the delta-component at $x=0$ (i.e., only wet days) and, therefore, the correlation between $M$ and $K$ decreases monotonically with increasing $n$. In contrast, when $q$ is close to $1$, there are large dry patches, following a wet day. During this dry patch, both $M$ and $K$ do not change. They tend to stay together at their current values during the long dry patch and this ``staying together'' makes them correlated. Hence, for large $q$, there are two competing correlation and decorrelation mechanisms. For smaller values of $n$ the correlation mechanism ``wins'', leading to an increase of $C_q(n)$, while for very large $n$, the decorrelation mechanism dominates, leading to an eventual decrease in $C_q(n)$. However, the fact that there is a sharp transition between these two behaviors, namely the monotonic and the non-monotonic is surprising and far from trivial. Hence, it is natural to wonder whether this shape transition shows up in real precipitation data. For this purpose we analyzed a large set of daily precipitation data from the German weather service (DWD)~\cite{dwd}. The data showed a quantitative agreement with our theoretical predictions and also showed the shape transition, as seen in Fig. \ref{fig:dwd} and discussed in detail later. Another interesting result concerns the decay of $C_q(n)$ for large $n$, that results from the decorrelation mechanism discussed above. We show that $C_q(n)$ decays as a power law $C_q(n) \sim n^{-\nu}$ for $n \gg 1$ (up to logarithmic corrections), where the exponent $\nu$ is independent of $q$, but depends on the tail of PDF $f(x)$. For example, if $f(x) \sim e^{-x^\gamma}$ for large $x$, with $\gamma >0$, then the exponent $\nu = 1$ is universal. In contrast, if $f(x) \sim 1/x^{\mu +1}$ with $\mu >1$, then the exponent $\nu = 1 - 1/\mu$ depends continuously on $\mu$.

We start by proving the independence of $K$ and $M$ in an IID sequence with a continuous $p(x)$. This corresponds to setting $q=0$ in Eq. (\ref{eq:mixed}), i.e., in the absence of dry-days, where $p(x) = f(x)$ with a continuous $f(x)$. We will thus express all our results in terms of the continuous $f(x)$. We just outline the derivation here and for details see the Supplemental Material (SM) \cite{SM}. Our main idea is to compute the joint probability distribution function (JPDF) $P(K,M\vert n)$ and show that it factorizes, i.e., $P(K,M\vert n) = P(K\vert n) \, P(M\vert n)$, indicating that they are statistically independent. It turns out to be difficult to derive this JPDF directly without keeping track of other features of the realisation of the time series. The main idea then is to first compute the ``grand JPDF'' of the record values $\vec{R} = \{R_1, R_2, \cdots, R_K = M\}$, the record ages $\vec{\ell} = \{\ell_1, \ell_2, \cdots, \ell_{K-1}, \ell \}$ and the number of records $K$ in the sequence (see Fig. \ref{fig:schematic}) and then integrate out $\{R_1, R_2, \cdots, R_{K-1}\}$ and $\vec{\ell}$, keeping $R_K = M$ and $K$ fixed. This then gives the marginal JPDF $P(K,M\vert n)$. It turns out that this grand JPDF $P(\vec{R}, \vec{\ell}, K \vert n)$ can be expressed in a closed form as shown in the SM \cite{SM}. Skipping details, we find that the generating function (GF) of $P(K,M\vert n)$ has a rather compact expression    
\begin{equation}\label{JPDF_GF}
\sum_{n=1}^\infty P(K,M\vert n)\, z^n = \frac{z f(M)}{1-zF(M)}  \frac{\left[-\ln\left(1-z F(M)\right) \right]^{K-1}}{(K-1)!} \;,
\end{equation}
valid for all $K \geq 1$, where $F(M) = \int_0^M f(x)\,dx$ is the cumulative distribution corresponding to $f(x)$. From Eq.~(\ref{JPDF_GF}), one can further sum over $K$ or integrate over $M$ to compute the GF of the marginal PDF $P(K \vert n)$ and~$P(M \vert n)$. From this, we compute the GF of the product, i.e., $\sum_{n=1}^\infty P(K \vert n) \,P(M \vert n) z^n$, which turns out to be identical to Eq. (\ref{JPDF_GF}) -- see \cite{SM}. This clearly shows the desired independence, i.e.,
\bea\label{eq_indep}   
P(K,M\vert n) = P(K\vert n) \, P(M\vert n) \;,
\eea
valid for all $n \geq 1$ and arbitrary {\it continuous} $f(x)$. This implies that the covariance in Eq. (\ref{eq:defCN}) also vanishes, i.e., $C_0(n) = 0$ for all $n \geq 1$.  This approach may look a bit algebraic but it turns out to play a crucial role in the derivation of the exact results for the dry-day model that we discuss below.

We now turn to the ``dry-day'' model with a nonzero $0<q< 1$ as in Eq. (\ref{eq:mixed}) and compute the JPDF $P_q(K,M \vert n)$. \blue{Throughout, records are counted among wet-day observations only: consequently $K=M=0$ if all $n$ days are dry.}
To proceed for $q>0$, we first define another random variable $n_W$ that represents the number of wet days in a given time series of size $n$. Clearly one has $0\leq n_W \leq n$. For a fixed $n$, since each entry is either dry with probability $q$ or wet with probability $(1-q)$, it follows that $n_W$ is simply distributed binomially
\begin{equation} \label{Binomial}
P(n_W \vert n) = {n \choose n_W} (1-q)^{n_W}\, q^{n-n_W} \quad, \quad 0 \leq n_W \leq n \;.
\end{equation}
For $q=0$, $P(n_W \vert n) = \delta_{n_W,n}$ and, in this case, $K$ and $M$ are completely uncorrelated as shown in Eq. (\ref{eq_indep}). For every wet day, the value of the entry $X_i >0$ is distributed via a continuous PDF $f(x)$ that is supported over $x \in [0, + \infty)$ \blue{[see Eq.~\eqref{eq:mixed}]}. 
In the context of precipitation time series, this makes perfect sense since the rainfall is necessarily nonnegative. We first make the crucial observation that, conditioned on $n_W$, the random variables $K$ and $M$ are completely independent, as shown above (\ref{eq_indep}). But, once averaged over $n_W$ drawn from Eq. (\ref{Binomial}), they do get correlated. This is reflected in  
\begin{equation} \label{PqKM}
P_q(K,M \vert n) = \sum_{n_W = 0}^n P_c(K,M \vert n_W) {n \choose n_W} (1-q)^{n_W}\, q^{n-n_W} 
\end{equation}
 where $P_c(K,M \vert n_W)$ is the conditional joint distribution of $K$ and $M$, given the number of nonzero entries $n_W$. Note that for $n_W>0$, we have
$P_c(K,M \vert n_W) = P(K \vert n_W) P(M \vert n_W)$ where $P(K \vert n_W)$ and $P(M \vert n_W)$ are the marginal distributions corresponding to the $q=0$ case discussed above and their explicit	GFs are given in \cite{SM}.   
However, when $n_W = 0$, we have $P_c(K,M \vert n_W = 0)= \delta_{K,0} \delta(M)$, since in this case there is no nonzero entry and, hence, we have necessarily $K=0$ and also $M=0$. It is useful to define the double GF with respect to $K$ and $n$, keeping $M$ fixed, as 
$\tilde P_q(u,M \vert v) = \sum_{n=0}^\infty v^n\,\sum_{K=0}^\infty u^K P_q(K,M \vert n)$. From Eq. (\ref{PqKM}) and using the expression of 
$P_c(K,M \vert n_W)$ discussed above, we find, after a few steps of algebra (see \cite{SM}) the following compact expression
\bea \label{def_GF3}
\tilde P_q(u,M \vert v) = \frac{\delta(M)}{1-q\,v} + \frac{\partial}{\partial M} \left[ \frac{(1-q v)^{u-1}}{(1- v g(M))^u}\right] \;. 
\eea
where $g(M)= q+(1-q) \int_0^M f(x) \,dx$. From (\ref{def_GF3}), one can derive the GFs of the marginal distributions of $K$ and $M$ \cite{SM}. These results lead to a compact expression for the double GF of the ``connected'' JPDF
$\left[P_q(K,M \vert n) - P_q(K\vert n) P_q(M \vert n)\right]$. From this, one can derive the GF of the covariance $C_q(n)$ in Eq. (\ref{eq:defCN}). The details are given in \cite{SM} and here we just quote our final result  
\begin{equation} \label{GF_Cq_final}
\sum_{n=0}^\infty C_q(n)\,v^n =\int_0^\infty \frac{dM}{1-v\,g(M)} \ln \left[ \frac{1-q \, v \, g(M)}{1-q\,v}\right] \;,
\end{equation}
where $g(M)$ is defined above. This is our main analytical result, valid for arbitrary $f(x)$ with finite first moment and $0\leq q\leq 1$. Clearly, the right hand side (RHS) vanishes as $q \to 0$, consistent with our previous result $C_0(n) = 0$ for all $n \geq 1$. {In fact, by expanding the RHS of Eq. (\ref{GF_Cq_final}) in powers of $v$ and identifying the coefficients of the same powers on both sides of (\ref{GF_Cq_final}) (see End Matter), one can compute $C_q(n)$ explicitly as 
\bea \label{explicit_Cq}
C_q(n) = \sum_{k=1}^n \frac{q^k}{k}\int_0^\infty [g(M)^{n-k} - g(M)^n] \, dM \;.
\eea
From these exact formulae (\ref{GF_Cq_final}) and (\ref{explicit_Cq}), we can derive the asymptotic behavior of $C_q(n)$ both for $n$ large, which is more conveniently done from (\ref{GF_Cq_final}) and $n$ small, which can be read off directly from (\ref{explicit_Cq}). For large $n$, we need to analyse the $v \to 1$ behavior of the GF in Eq. (\ref{GF_Cq_final}). This gives the large $n$ asymptotic power law decay of $C_q(n) \sim n^{-\nu}$, as discussed in the results section before -- the details of the computations can be found in \cite{SM}. On the other hand, from Eq.~(\ref{explicit_Cq}), one gets immediately $C_q(n)$ for $n=1,2,3,\cdots$ in terms of $f(x)$.} In particular, by comparing $C_q(1)$ and $C_q(2)$, we find (see End Matter) that $C_q(2) > C_q(1)$ if $q>q_c$, while, for $q<q_c$, we get $C_q(2)< C_q(1)$. This critical value of $q_c$ is given explicitly by (see End Matter)
\begin{equation} \label{cond_qc}
q_c =\frac{1}{2} \left(- b + \sqrt{b^2+8}\right)~~ {\rm with} ~~ b = \left(1+{2\mu_1}/{\mu_2}\right)\;,
\end{equation}
where $\mu_1 = \int_0^\infty \bar{F}(M)\,dM$ and $\mu_2 = \int_0^\infty \bar{F}^2(M)\,dM$, where $\bar{F}(M) = \int_M^\infty f(x)\,dx$. For example, for the purely exponential PDF $f(x) = e^{-x}$, we get $q_c = (\sqrt{33}-5)/2 = 0.372281\ldots$ and for the uniform distribution $f(x) = 1$ for $x \in [0,1]$, we get $q_c = \sqrt{6}-2 = 0.44949 \ldots$. In Fig. \ref{Fig_Exp}, we compare our theoretical predictions for the exponential case with direct simulations, finding excellent agreement. In particular the transition at $q=q_c = 0.372281\ldots$ is clearly visible.

An interesting scaling limit is when $q \to 1$ (very few wet days) and $n \to \infty$, keeping the product $t = n(1-q)$ fixed. 
In this limit, the rare wet days form a continuous-time Poisson process where the $t = n(1-q)$ plays the role of time. In the literature on record statistics, this limiting process is known as ``the random record model''~\cite{arnold,Westcott77,BN92,DGP08,PK08}. In this case, by taking the scaling limit in Eq. (\ref{GF_Cq_final}), 
we find that $C_q(n)$ approaches a scaling form 
$C_q(n) \approx C(t = n\,(1-q))$ where $C(t)$ can be obtained explicitly in terms of $\bar F(M)$ (see End Matter). The function $C(t)$ has a non-monotonic behavior with $t$, since the $q \to 1$ limit implies $q>q_c$. The asymptotic behaviours of $C(t)$ can also be extracted explicitly. For small $t$, the scaling function $C(t) \approx \mu_1\,t$ displays a universal linear behavior, while for large $t$, it decays as a power law with an exponent $\nu$ (as in the discrete model).

\noindent \emph{Comparison to real precipitation data.} The shape transition is a prediction without adjustable parameters that can be tested directly on rainfall records, where dry days provide a natural delta peak at $x=0$. We used daily precipitation data from 6,105 stations of the German weather service (DWD)~\cite{dwd}. To suppress day-to-day correlations, we skip $l=12$ days between consecutive entries; the End Matter shows that the measured covariance then converges to the IID prediction. Because the dry-day fraction varies strongly from station to station, we pooled stations with similar $q$. Set A consists of the 232 stations that have complete daily records over 1974--2013 and $q\in[0.48,0.52]$, giving $q\approx0.5016$. From the empirical wet-day distribution $f(x)$ (shown in~\cite{SM}) we compute $\mu_1$ and $\mu_2$, and Eq.~\eqref{cond_qc} predicts $q_c\approx0.3152$. Since $q>q_c$, the theory predicts a nonmonotonic $C_q(n)$. To probe the opposite regime, we built set B from the November records of 15 stations with a November dry-day fraction of at most 0.275, giving $q\approx0.2323$ and, by the same procedure, $q_c\approx0.2795$ (here $l=10$). Since now $q<q_c$, the theory predicts a monotonic decay. In both cases we obtain the theoretical $C_q(n)$ from Eq.~\eqref{GF_Cq_final} with the measured $q$ and $f(x)$, without any adjustable parameter, and compare it with $C_q(n)$ measured directly from the same records. Figure~\ref{fig:dwd} shows excellent agreement in both regimes: the data display the nonmonotonic shape for set A and the monotonic decay for set B, exactly as predicted from the respective values of $q$ and $q_c$.

\noindent  \emph{Conclusion.} 
%
Our results show that the exact independence of the maximum $M$ and the number of records $K$, which holds for all $n$ in IID series drawn from a continuous distribution, is fragile: a delta peak at $x=0$ in $p(x)$ generates correlations for any $0<q<1$. In the dry-day model of Eq.~(\ref{eq:mixed}), $C_q(n)$ undergoes a sharp shape transition at an exactly computable $q_c$, from monotonic decay for $q<q_c$ to a nonmonotonic behavior, rising before it decays, for $q>q_c$. In both regimes, $C_q(n)$ decays algebraically at large $n$ (up to logarithmic corrections), with an exponent independent of $q$ and set by the tail of $f(x)$. Both regimes are observed in German precipitation data, in quantitative agreement with the theory.

A natural question is how the covariance $C_q(n)$ behaves for time series with correlated entries, going beyond the IID case. Indeed, in the real precipitation time series discussed above, the entries $X_i$'s happen to be weakly correlated with a finite correlation time. This can be modelled by the so-called AR(1) process, that appears in several applications ranging from climate, finance to queuing theory \cite{Book_OUP,BD02,Lar04}. Here the entries can be expressed by a stochastic recursion relation $X_{i+1}=aX_i+\eta_i$, where $0\leq a \leq 1$ and the 
$\eta_i$'s are IID noises. For $a=0$ it reduces to the IID model discussed in this Letter but for finite $a>0$, the entries $X_i$'s are correlated over a time scale $\tau=1/|\ln a|$. Our preliminary studies \cite{us_inprogress} on the covariance between $M$ and $K$ show that it always behaves non-monotonically as a function of $n$, for all $a > 0$. It would be interesting to obtain the covariance exactly in this correlated time series 
in the presence of dry days, i.e., with a nonzero $q$.

\begin{acknowledgments}
We acknowledge support from ANR
Grant No. ANR-23-CE30-0020-01 EDIPS. Part of this work was completed during a visit at 
the Kavli Institute for Theoretical Physics (KITP) supported by grant NSF PHY-2309135.
\end{acknowledgments}

\clearpage


\setcounter{equation}{0}
\setcounter{figure}{0}
\renewcommand{\theequation}{EM\arabic{equation}}
\renewcommand{\thefigure}{EM\arabic{figure}}

\section{End matter}

\subsection{Illustration of the dry-day model in Eq. (\ref{eq:mixed})}

\begin{figure}[ht]
\includegraphics[width = \linewidth]{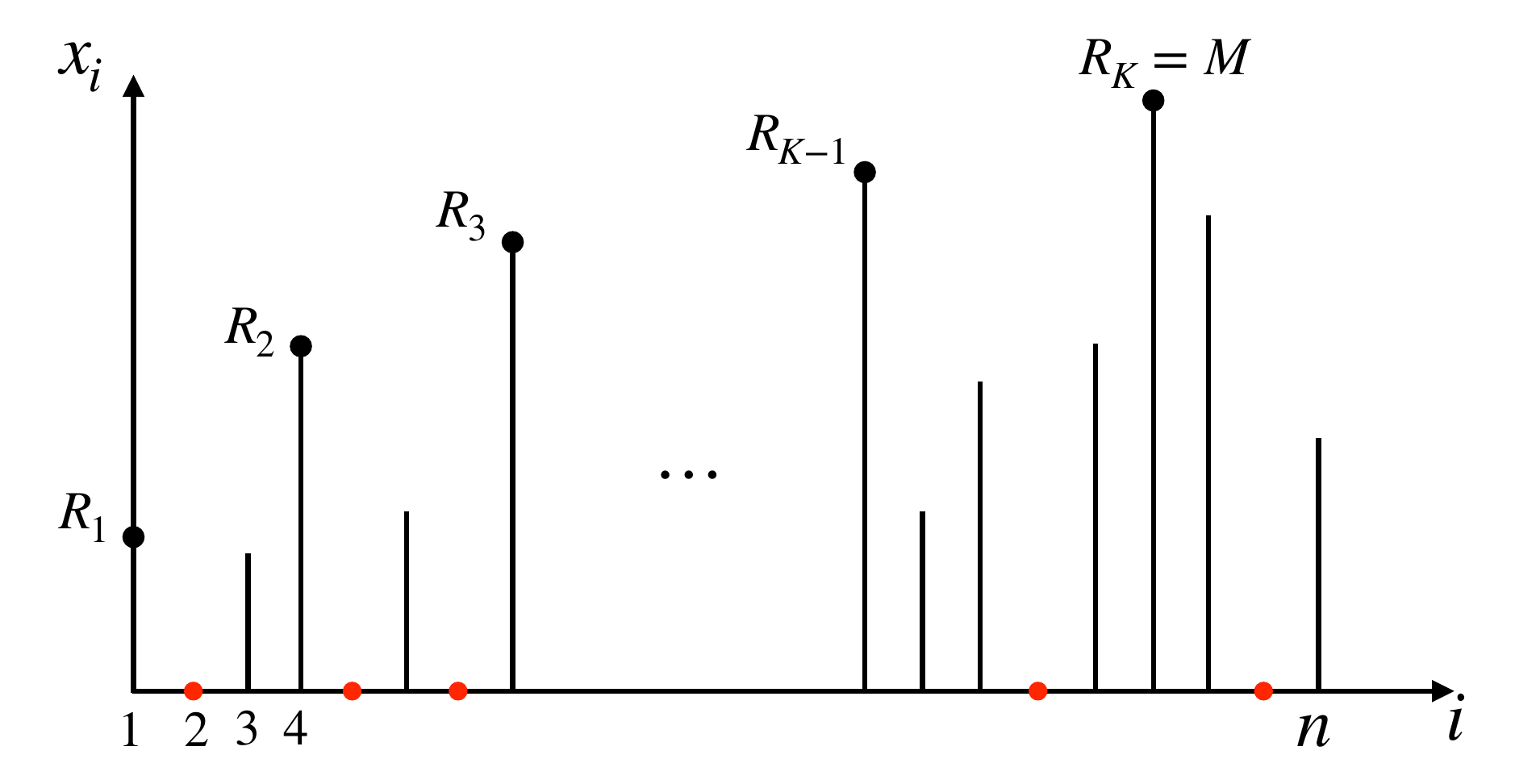}
\caption{{A typical realization of an IID time series of $n$ entries and $K$ records in the dry-day model. Each entry is either zero (dry day), with probability $q$, indicated by a red dot, or drawn from a continuous distribution $f(x)$ (wet day), with probability $1-q$ [see Eq.~(\ref{eq:mixed}) in the Letter]. The record values $R_1,R_2,\ldots,R_K$ are indicated by black dots, with the last record $R_K=M$ corresponding to the global maximum.}}\label{Fig_intro_dry}
\end{figure}

\subsection{Inverting the generating function of $C_q(n)$}

To invert the generating function in (\ref{GF_Cq_final}), one first expands the two factors in the integrand in powers of $v$. This gives
\bea 
\hspace*{-0.8cm}&& \frac{1}{1-v\,g(M)} =  \sum_{k=0}^\infty v^k [g(M)]^k \;, \label{expansion_v1}\\
\hspace*{-0.8cm} &&\ln \left[ \frac{1-q \, v \, g(M)}{1-q\,v}\right] = \sum_{k=1} ^\infty \frac{q^k v^k}{k} \left( 1- [g(M)]^k\right) \;. \label{expansion_v2}
\eea
It then follows that the product of the two factors admits the expansion
\bea 
&&\frac{1}{1-v\,g(M)} \;\ln \left[ \frac{1-q \, v \, g(M)}{1-q\,v}\right]  \label{GF_product} \\
&&= \sum_{n=0}^\infty v^n \sum_{k=1}^n \frac{q^k}{k} \left( [g(M)]^{n-k} - [g(M)]^n\right) \;.\nonumber
\eea
Substituting this power series expansion in the RHS of Eq.~(\ref{GF_Cq_final}) and identifying the powers of $v$ on both sides lead to the result in (\ref{explicit_Cq}).

\subsection{The covariance $C_q(n)$ for small $n$}

We provide the exact formulas for $C_q(1)$ and $C_q(2)$ obtained by expanding the RHS of Eq. (\ref{GF_Cq_final}) up to order $O(v^2)$ for small $v$. One gets, from the $O(v)$ term,  
\begin{equation} \label{Cq1}
C_q(1) = q(1-q) \, \mu_1 \quad, \quad {\rm where} \quad  \mu_1 = \int_0^\infty \bar{F}(M)\, dM \;,
\end{equation}
where $\bar{F}(M) = \int_M^\infty f(x)\,dx$

Similarly, comparing the $O(v^2)$ term on both sides of Eq. (\ref{GF_Cq_final}) gives
\bea \label{Cq_2_1}
C_q(2) &=& q \int_0^\infty dM \, g(M)(1-g(M)) \nonumber \\
&+& \frac{q^2}{2} \int_0^\infty dM\, (1 - g^2(M)) \;.
\eea
Substituting $g(M) = 1 -(1-q)\, \bar{F}(M)$ and simplifying one gets
\begin{equation} \label{Cq_2_2}
C_q(2) = q(1-q) \left[ (1+q)\, \mu_1 - (1-q)\,\left(1 +\frac{q}{2} \right)\,\mu_2\right] \;,
\end{equation}
where $\mu_2 = \int_0^\infty \bar{F}^2(M)\, dM$. Note that these results are valid for a parent distribution $f(x)$ that has an infinite support over the full positive real line. For a parent distribution with a finite support over $[0,x^*]$, the same formulae as in Eqs. (\ref{Cq1}) and (\ref{Cq_2_2}) will hold but with the upper limit $+\infty$ in the definition of $\mu_1$ and $\mu_2$ replaced by $x^*$. This yields
\begin{equation} \label{Cq_diff}
C_q(2) - C_q(1) = q(1-q) \left[q\,\mu_1 - (1-q)\left(1+\frac{q}{2} \right)\, \mu_2 \right] \;.
\end{equation}
Thus, as $n$ increases from $n=1$ to $n=2$, the correlator will either decrease or increase, depending on the sign of the RHS of Eq. (\ref{Cq_diff}). It will increase if 
\bea \label{cond_inc}
q\,\mu_1 - (1-q)\left(1+\frac{q}{2} \right)\, \mu_2 > 0 \;,
\eea
which is equivalent to having
\bea \label{cond_inc2}
q^2 + b\,q-2 > 0 \quad, \quad b = \left(1 + \frac{2 \mu_1}{\mu_2} \right) \;.
\eea
The LHS can be written as $(q-q_+)(q-q_-)$ where 
\bea \label{qpm}
q_{\pm} = \frac{1}{2} \left(- b \pm \sqrt{b^2+8}\right) \;.
\eea
Thus the condition in Eq. (\ref{cond_inc2}) reads
\bea \label{cond_inc3}
(q - q_+)(q-q_-) > 0  \;.
\eea
Since $q_-<0$, it follows that this condition is satisfied provided 
\begin{equation} \label{cond_qc_em}
q>q_c = q_+ =\frac{1}{2} \left(- b + \sqrt{b^2+8}\right) \;,
\end{equation}
which is stated in Eq. (\ref{cond_qc}) in the main text.

\subsection{The random record model}

\begin{figure}[t!]
\includegraphics[width=0.9\linewidth]{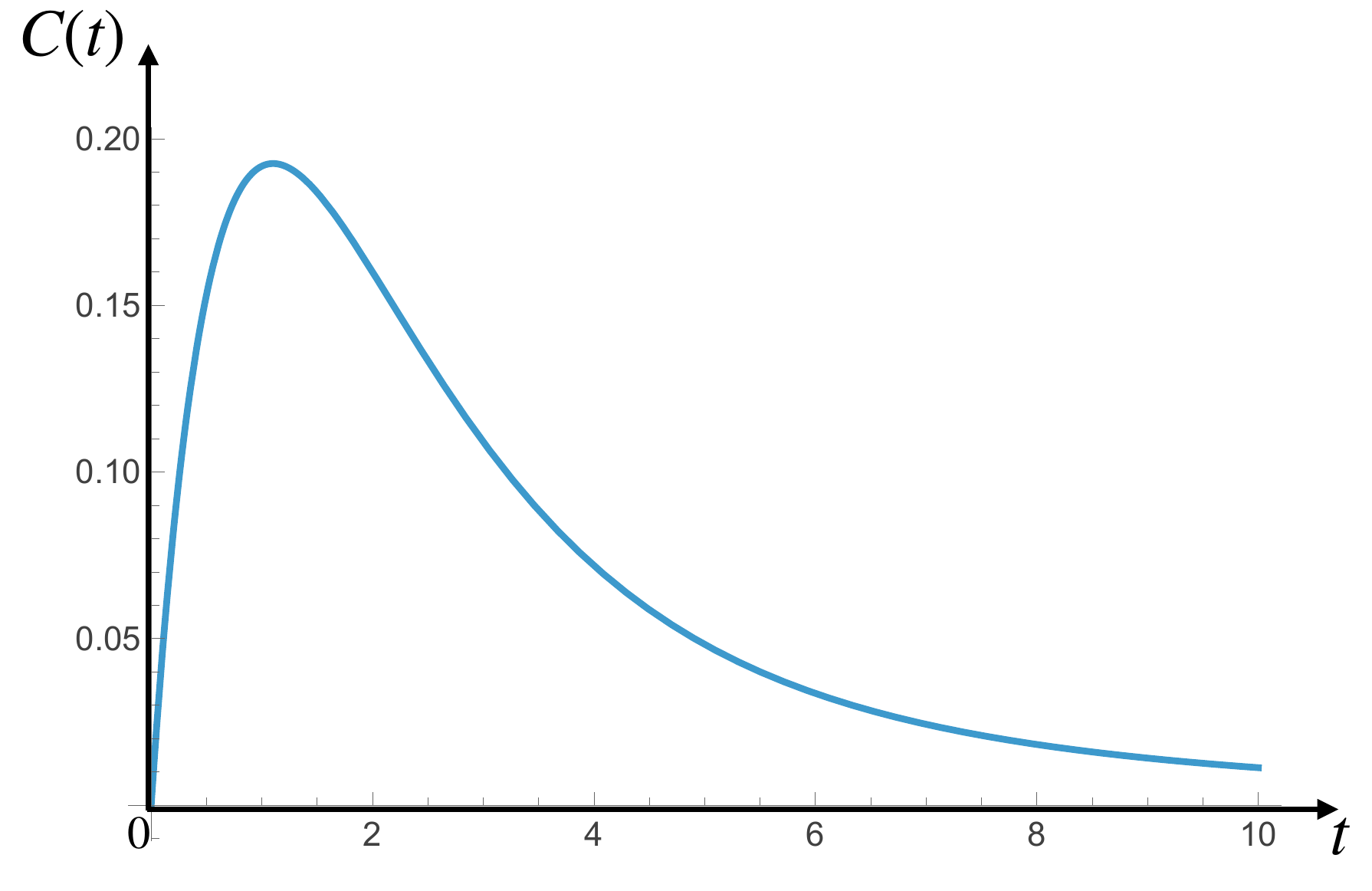}
\caption{A plot of $C(t)$ in Eq. (\ref{C_cont_unif4}) vs $t$.}\label{Fig_Ct_exp}
\end{figure}

As discussed in the main Letter, this corresponds to the scaling limit where $q \to 1$ (very few wet days) and $n \to \infty$, keeping the product $t = n(1-q)$ fixed. In this limit, one can show \cite{SM} that $C_q(n) \approx C(t = (1-q)\,n)$ where, using the expression for generating function in (\ref{GF_Cq_final}), the Laplace transform of $C(t)$ is given exactly by
\be \label{Laplace_C}
\int_0^\infty C(t)\, e^{-st} \, dt = \int_0^\infty \frac{dM}{s + \bar{F}(M)} \ln \left(1+ \frac{\bar F(M)}{s+1} \right) \;,
\ee
where we recall that $\bar{F}(M) = \int_M^\infty f(x)\,dx$. Note that this result holds for a parent distribution $f(x)$ that is supported over the full positive line. For $f(x)$ with a finite support $[0,x^*]$, it is easy to see that the same result as in Eq. (\ref{Laplace_C}) holds with the upper limit $+\infty$ replaced by $x^*$. In fact, the Laplace transform in Eq. (\ref{Laplace_C}) can be explicitly inverted for any $\bar{F}(M)$, leading to
\be
C(t)
=
\int_0^\infty dM\,
e^{-t\bar F(M)}
\left[
{\cal H}(t)
-
{\cal H}\left((1-\bar F(M))t\right)
\right] \;,
\label{Ct_H}
\ee
where we introduced the function
\be
{\cal H}(x)
=
\int_0^x \frac{1-e^{-u}}{u}\,du
=
\gamma_{\rm E}+\ln x+E_1(x) \;.
\label{Hdef}
\ee
Here 
\be
E_1(x)=\int_x^\infty \frac{e^{-u}}{u}\,du,
\qquad x>0,
\label{E1def}
\ee
denotes the exponential integral function. In particular, for the uniform distribution $f(x) = 1$ for $x \in [0,1]$ and $f(x)=0$ elsewhere, giving $\bar{F}(M)=1-M$ for $0<M<1$. 
The RHS in (\ref{Ct_H}), in this case, can be evaluated in a closed form leading to 
\begin{equation} 
\label{C_cont_unif4}
C(t) = \frac{2}{t}\,e^{-t}\int_0^t \frac{(\cosh x - 1)}{x}\,dx \quad, \quad t \geq 0 \;.
\end{equation}

\vspace*{0.3cm}
\noindent A plot of this function $C(t)$ is shown in Fig.~\ref{Fig_Ct_exp}. It has the following asymptotic behaviours
\bea \label{Ct_asympt}
C(t) \approx
\begin{cases}
&\dfrac{t}{2} - \dfrac{t^2}{2} + O(t^3) \quad, \quad t \to 0 \;, \\
& \\
& \dfrac{1}{t^2} + O(t^{-3} )\quad, \quad \quad\, t \to +\infty \;.
\end{cases}
\eea

\begin{figure}[b!]
    \centering
    \includegraphics[width=0.95\linewidth]{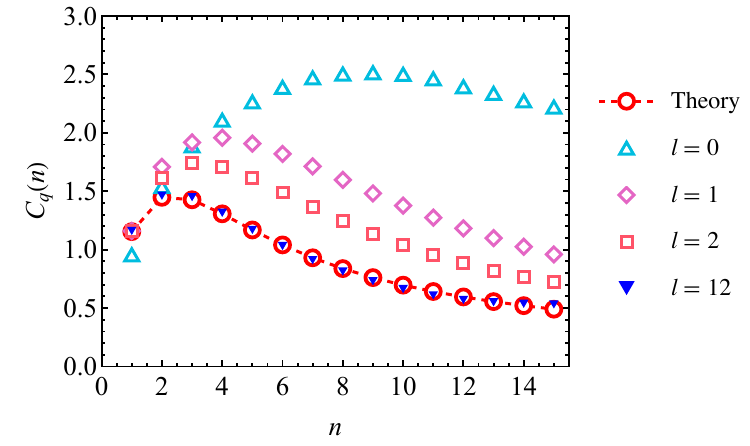}
    \caption{The covariance $C_q(n)$ computed from the rainfall data over the 40-year period 1974--2013 for which $q\in [0.48, 0.52]$. The different graphs are obtained by skipping $l$ days between two successive data. It can be seen that for $l=12$, the observed data fit the theoretical result for IID case.  }
    \label{fig:cqnwithl}
\end{figure}
\subsection{Convergence of the precipitation data to the dry-day model}

In Fig.~\ref{fig:dwd} (left) in the main text, we have skipped $l = 12$ days between consecutive
observations to suppress temporal correlations in the precipitation data.
The resulting statistics agree well with the dry-day model prediction for IID
random variables. Figure~\ref{fig:cqnwithl} shows that, as $l$ increases from $l=0$ to $l=12$,
the covariance $C_q(n)$ computed from the observational data converges to
the IID result.   


\begin{thebibliography}{9}



\bibitem{Gumbel58}
E. J. Gumbel, {\it Statistics of Extremes}, (Dover, New York, 1958).

\bibitem{Lead82}
M.~R. Leadbetter, G. Lindgren, and H. Rootz{\'e}n, {\it Extremes and related properties of random sequences and
processes}, (Springer-Verlag, New York, 1982).

\bibitem{arnold}
B. C. Arnold, N. Balakrishnan, and H. N. Nagaraja,
\emph{Records} (Wiley, New York, 1998).

\bibitem{Resnick08}
S. I. Resnick, {\it Extreme values, regular variation, and point processes}, Vol. 4. Springer Science \& Business Media (2008).

\bibitem{Review_EVS}
S. N. Majumdar, A. Pal, G. Schehr, {\it Extreme value statistics of correlated random variables: a pedagogical review}, Phys. Rep. {\bf 840}, 1 (2020).


\bibitem{Book_OUP}
S. N. Majumdar, G. Schehr, {\it Statistics of Extremes and Records in Random Sequences}, Oxford University Press (2024).

\bibitem{Fis25}
E. M. Fischer, M. Bador, R. Huser, E. J. Kendon, A. Robinson, S. Sippel, {\it Record-breaking extremes in a warming climate}, Nat. Rev. Earth Environ. {\bf 6}, 456 (2025).


\bibitem{SK99}
B. Schmittmann, R. K. P. Zia, {\it “Weather” records: Musings on cold days after a long hot Indian summer},
Am. J. Phys., {\bf 67}, 1269 (1999).

\bibitem{Easter}
D. R. Easterling, G. A. Meehl, C. Parmesan, S. A. Changnon, T. R. Karl, L. O. Mearns, {\it Climate extremes: observations, modeling, and impacts},  Science {\bf 289}, 2068 (2000).

\bibitem{Redner06}
S. Redner, M. R. Petersen, {\it Role of global warming on the statistics of record-breaking temperatures}, Phys. Rev. E {\bf 74}, 061114 (2006).

\bibitem{WK10}
G. Wergen, J. Krug, {\it Record-breaking temperatures reveal a warming climate}, EPL {\bf 92}, 30008 (2010).

\bibitem{Hansen2012}
J. Hansen, M. Sato, R. Ruedy, {\it Perception of climate change}, Proc. Nat. Acad. Sci. {\bf 109}, E2415 (2012).

\bibitem{rainfall}
E. M. Fischer, R. Knutti, {\it Anthropogenic contribution to global occurrence of heavy-precipitation and high-temperature extremes}, Nat. Clim. Change {\bf 5}, 560 (2015).

\bibitem{drought}
B. I. Cook, T. R. Ault, J. E. Smerdon, {\it Unprecedented 21st century drought risk in the American Southwest and Central Plains}, Sci. Adv. {\bf 1}, e1400082 (2015).










\bibitem{record-review}
G. Wergen, \emph{Records in stochastic processes -- theory and applications},
J. Phys. A \textbf{46}, 223001 (2013).


\bibitem{GMS17}
C. Godr{\`e}che, S. N. Majumdar, G. Schehr, {\it Record statistics of a strongly correlated time series: random walks and L{\'e}vy flights}, J. Phys. A: Math. Theor. {\bf 50}, 333001 (2017).


\bibitem{Westcott77} M. Westcott, {\it The random record model},  Proc. R. Soc. A {\bf 356},  529 (1977).

\bibitem{BN92} J. A. Bunge and H. N. Nagaraja, {\it Dependence structure of Poisson-paced records}, J. Appl. Prob. {\bf 29}, 587 (1992).



\bibitem{DGP08} J. Davidsen, P. Grassberger, and M. Paczuski, {\it Networks of recurrent events, a theory of records, and an application to finding causal signatures in seismicity}, Phys. Rev. E {\bf 77}, 066104 (2008).

\bibitem{PK08} S.-C. Park and J. Krug, {\it Evolution in random fitness landscapes: the infinite sites model}, J. Stat. Mech.: Theory Exp. P04014 (2008).


\bibitem{MBK19}
S. N. Majumdar, P. von Bomhard and J. Krug, {\it Exactly solvable record model for rainfall}, Phys. Rev. Lett. {\bf 122}, 158702 (2019).


\bibitem{dwd}
Deutscher Wetterdienst (DWD), Climate Data Center (CDC), \texttt{https://www.dwd.de}.

\bibitem{SM}
S. N. Majumdar, S. Sabhapandit, G. Schehr, see Supplementary Material.

\bibitem{us_inprogress}
S. N. Majumdar, S. Sabhapandit, G. Schehr, in preparation.

\bibitem{BD02}
\blue{P. J. Brockwell, R. A. Davis, {\it Introduction to time series and forecasting},  New York, NY (Springer New York) (2002).}

\bibitem{Lar04}
\blue{
H. Larralde, {\it Statistical properties of a discrete version of the Ornstein-Uhlenbeck process}, Phys. Rev. E {\bf  69}, 027102 (2004).}

\end{thebibliography}
\end{document}


\setcounter{secnumdepth}{3}
\renewcommand{\theequation}{SM.\arabic{equation}}

\title{Supplemental Material for\\[3pt] Emergence of Correlations Between Records and Maximum in Random Time Series}

\author{Satya N. Majumdar}
\affiliation{Universit\'e Paris-Saclay, CNRS, LPTMS, 91405 Orsay, France}

\author{Sanjib Sabhapandit}
\affiliation{Raman Research Institute, Bengaluru 560080, India}

\author{Gr\'egory Schehr}
\affiliation{Laboratoire de Physique Th\'eorique et Hautes \'Energies, Sorbonne Universit\'e and CNRS, 4 place Jussieu, 75252 Paris Cedex 05, France}

\date{\today}
\begin{abstract}
In this Supplemental Material we present the details leading to the main results presented in the Letter.
\end{abstract}
\maketitle

\tableofcontents
\clearpage


\begin{figure}[ht]
\includegraphics[width = 0.6\linewidth]{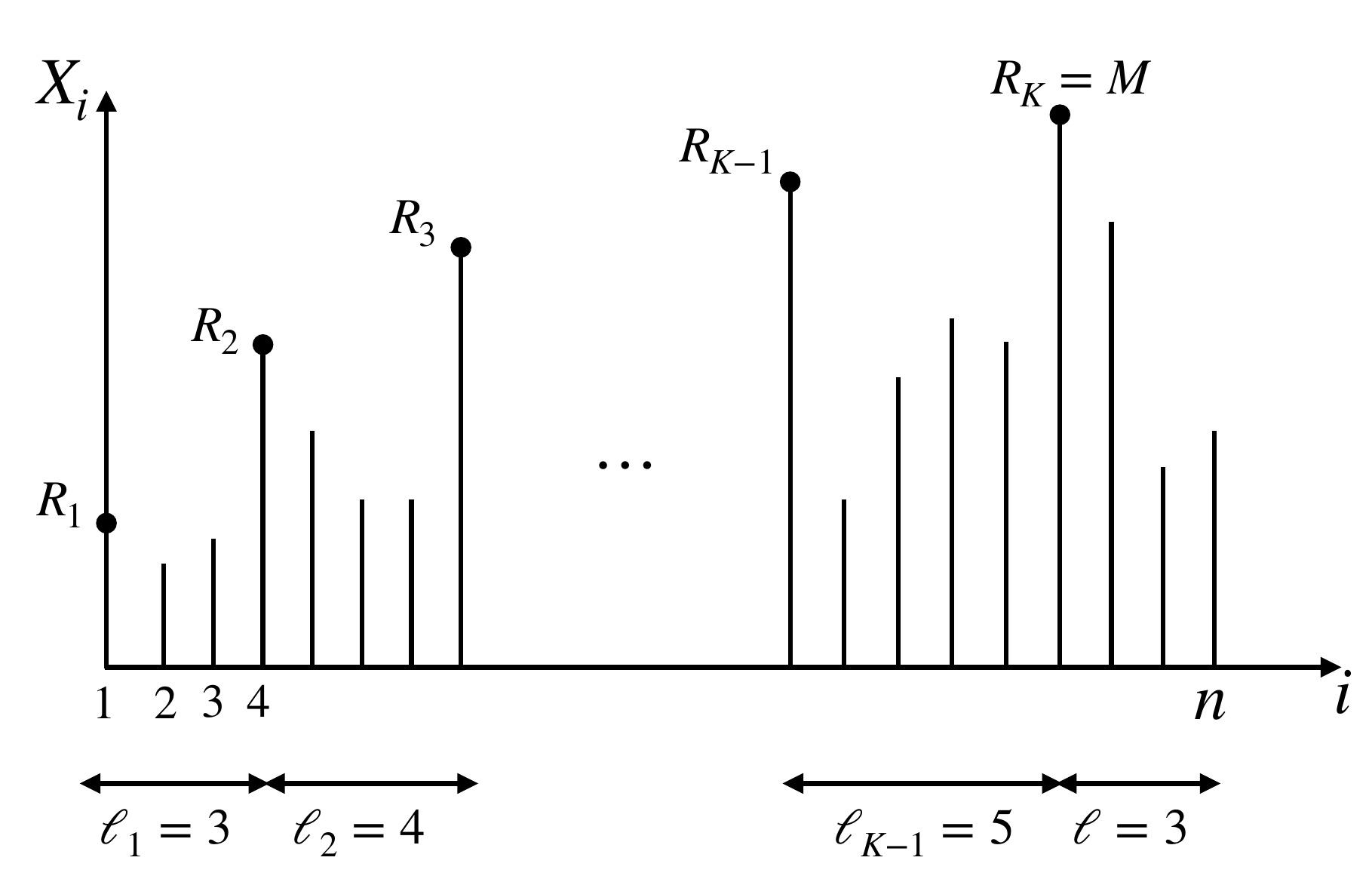}
\caption{A typical sequence of IID discrete time series with $n$ entries and $K$ records. Here $R_i$'s denote the record values, with $R_K = M$ denoting the global maximum, which is also the last record value. The $\ell_i$'s denote the record ages, i.e., the interval between successive records. The interval following the last record is denoted by $\ell$.}\label{Fig_intro}
\end{figure}

\section{Joint statistics of the number of records and the maximum for an IID sequence of size $n$}\label{sec:IID}

We consider a sequence of $n$ independent and identically distributed (IID) random variables $\{X_1, X_2, \cdots, X_n\}$ where each $X_i$ is drawn from a distribution $f(x)$ which is continuous and is supported over $x \in [0, +\infty)$. For example, it may represent a daily precipitation time series where $X_i$ represents the amount of rainfall in a given area on day $i$. In this sequence a record occurs at step $m$ if $X_m > \{X_1,X_2, \cdots, X_{m-1}\}$. Let us consider a sequence where the number of records is $K$, with  $R_i$ ($i=1,2,\cdots, K$) denoting the values of the records. {Clearly $K$ is a random variable.} Let $M$ denote the global maximum, i.e., $M = \max\{x_1, x_2, \cdots, x_n\}$. Clearly it follows that $M = R_K$ since the value of the last record in this sequence is necessarily the global maximum. {The maximum $M$ is random as well.} We are interested in the question whether the global maximum $M$ and the number of records $K$ are correlated. For this, we would like to compute the joint distribution $P(K,M\vert n)$, which will encode the correlations, if any. {Surprisingly, we will see that this joint distribution factorizes, i.e.,} 
\bea \label{joint_factor}
P(K,M\vert n) = P(K \vert n) P(M \vert n)  \;,
\eea
where $P(K \vert n)$ and $P(M \vert n)$ denote the marginal distributions of $K$ and $M$ respectively. This result is universal and holds 
for any underlying distribution $f(x)$ as long as it is continuous. Actually, this result is even more general and holds for any continuous $f(x)$, not necessarily supported on the positive side.

{It turns out that it is not easy to directly write down the evolution equation} for $P(K,M\vert n)$ with increasing $n$, even though it is possible to do the same for each of the marginal distributions (see e.g., \cite{Book_OUP}). To compute $P(K,M\vert n)$, we need to consider an enlarged set of degrees of freedom characterizing the configuration of $n$ variables with $K$ records. These degrees of freedom are (see Figure. \ref{Fig_intro}) (i) the record values $\{R_1, R_2, \cdots, R_K\}$ and (ii) the intervals $\{\ell_1, \ell_2, \cdots, \ell_{k-1}, \ell \}$
between two successive records. Note that the last interval denoted by $\ell$ specifies the number of entries after the last record. Note that 
$\{\ell_1, \ell_2, \cdots, \ell_{k-1}\}$ each ranges from $1$ to $+\infty$, while the last interval $\ell$ ranges from $0$ to $+\infty$. For compactness, we will denote these two sets of variables by $\vec{R} \equiv \{R_1, R_2, \cdots, R_K\} $ and $\vec{\ell} \equiv \{\ell_1, \ell_2, \cdots, \ell_{k-1}, \ell\}$. We will now show that the joint distribution 
$P(\vec{R}, \vec{\ell},K \vert n)$ of this enlarged set of variables can be written down explicitly and the idea will be then to integrate over $\{R_1, R_2, \cdots, R_{K-1}\}$ and sum over $\vec{\ell}$ to get the joint distribution $P(K,R_K=M\vert n)$. Below we carry out this procedure.

The joint distribution $P(\vec{R}, \vec{\ell},K \vert n)$ can be explicitly written as
\begin{equation} \label{JPDF_1}
P(\vec{R}, \vec{\ell},K \vert n) = \left[\prod_{i=1}^{K-1} f(R_i) \left[\int_0^{R_i} f(x)\, dx \right]^{\ell_i-1} \theta(R_{i+1}-R_i)\right] \; f(R_K) \left[\int_0^{R_K} f(x)\, dx \right]^{\ell}\,\delta\left( \sum_{i=1}^{K-1}\ell_i + \ell + 1 - n \right) \;,
\end{equation}
where $\theta(z)$ is the Heaviside step function, i.e., $\theta(z) = 1$ for $z>0$ and $\theta(z) = 0$ for $z<0$. This result can be best understood by following Fig. \ref{Fig_intro}. If $R_i$ and $R_{i+1}$ denote two successive record values, all the $\ell_{i}-1$ entries separating them must have values less than $R_i$. The probability of this block given $R_i$ is simply $\left[\int_0^{R_i} f(x) \,dx\right]^{\ell_{i}-1}$. Multiplying this by $f(R_i)$ and taking a product over all $(K-1)$ blocks (with the constraint $R_{i+1}>R_i$) gives the first term inside the parenthesis $\left[ \cdots \right]$ on the right hand side (RHS) of Eq. (\ref{JPDF_1}). The $\ell$ entries after the last record with value $R_K$ must all be less than $R_K$. The probability of this event is simply $f(R_K) \left[\int_0^{R_K} f(x)\,dx \right]^\ell$. This explains the second term between the parenthesis and the delta function in Eq. (\ref{JPDF_1}). Finally, the total number of entries is fixed to be $n$, indicating the hard constraint that $\ell_1 + \ell_2 + \cdots + \ell_{K-1} + \ell + 1 = n$. This hard constraint is encoded by the delta function in Eq. (\ref{JPDF_1}). Taking the generating function with respect to $n$ (that gets rid of the delta function) and summing over $\vec{\ell}$ gives a rather simplified expression
\bea \label{JPDF_GF_1}
\sum_{\vec \ell} \sum_{n=1}^\infty  P(\vec{R}, \vec{\ell},K \vert n)\,z^n =  \frac{z\, f(R_K)}{1- z F(R_k)} \prod_{i=1}^{K-1} \frac{z\, f(R_i)}{1 - z F(R_i)} \theta(R_{i+1}-R_i) \;,
\eea
where we have defined the cumulative distribution 
\bea \label{def_F}
F(R) = \int_0^R f(x)\,dx \;.
\eea
We now fix $R_K = M$ and integrate over $\{R_1, R_2, \cdots, R_{K-1} \}$. This gives the generating function of our desired joint distribution $P(K,M \vert n)$
\bea \label{JPDF_GF_2}
\sum_{n=1}^\infty P(K,M\vert n)\, z^n = \frac{z f(M)}{1-zF(M)} I_{K-1}(M) \;,
\eea 
where $I_{K-1}(M)$ is a nested integral
\bea \label{def_IK}
I_{K-1}(M) = \int_0^M \frac{z f(R_{K-1})}{1-z F(R_{K-1})}\, dR_{K-1} \int_{0}^{R_{K-1}} \frac{z f(R_{K-2})}{1-z F(R_{K-2})}\, dR_{K-2} \cdots  \int_{0}^{R_{2}} \frac{z f(R_{1})}{1-z F(R_{1})}\, dR_{1} \;.
\eea
Differentiating with respect to (wrt) $M$ yields the recursion relation
\bea \label{recursion}
\frac{d I_{K-1}(M)}{dM} = \frac{z\,f(M)}{1-z F(M)}\, I_{K-2}(M) \;,
\eea
starting from $I_0(M) = 1$ for all $M \geq 0$. The exact solution of this recursion (\ref{recursion}) is given by (as can be easily proved by the method of induction)
\bea \label{IK}
I_{K-1}(M) = \frac{\left[-\ln\left(1-z F(M)\right) \right]^{K-1}}{(K-1)!} \;.
\eea
Hence we get the generating function of the joint distribution explicitly as
\bea\label{JPDF_GF_3}
\sum_{n=1}^\infty P(K,M\vert n)\, z^n = \frac{z f(M)}{1-zF(M)}  \frac{\left[-\ln\left(1-z F(M)\right) \right]^{K-1}}{(K-1)!} \;,
\eea
valid for all $K \geq 1$. Next, we compute the generating functions for the marginal distributions of $K$ and $M$ respectively. 

\vspace*{0.5cm}
\noindent{\bf Marginal distributions of $K$.} Let us first consider the marginal distributions of $K$. Integrating (\ref{JPDF_GF_3}) over $M$ gives
\bea \label{marg_K_1}
\sum_{n=1}^\infty P(K\vert n)\, z^n = \sum_{n=1}^\infty \int_{0}^\infty dM\, P(K,M\vert, n) z^n=  \int_0^\infty dM \, \frac{z f(M)}{1-zF(M)}  \frac{\left[-\ln\left(1-z F(M)\right) \right]^{K-1}}{(K-1)!} \;.
\eea
Performing the change of variable $u = - \ln(1 - z\,F(M))$, one reproduces the well know result (see e.g. \cite{Book_OUP})
\bea \label{marg_K_2}
\sum_{n=1}^\infty P(K\vert n)\, z^n  = \int_0^{-\ln(1-z)} du\, \frac{u^{K-1}}{(K-1)!} = \frac{\left[ - \ln(1-z)\right]^K}{K!} \quad, \quad K \geq 1 \;.
\eea 
In fact, expanding the RHS in powers of $z$ and matching, one can express $P(K \vert n)$ as (see e.g.~\cite{GMS17})
\bea \label{stirling}
P(K \vert n) = \dfrac{{n \brack K}}{n!} \;,
\eea
where ${n \brack K}$ denotes the Stirling numbers representing the number of permutations of $n$ with $K$ disjoint cycles.

 For future purposes,
it is useful to provide another alternative representation of $P(K\vert n)$. Multiplying Eq. (\ref{marg_K_2}) by $u^K$ and summing over
$K=1,2\ldots$, we get
\bea
\label{marg_K_alt.1}
\sum_{n=1}^{\infty}\sum_{K=1}^{\infty} P(K\vert n)\, u^K\, z^n = e^{-u\, \ln(1-z)}-1= \frac{1}{(1-z)^u}-1 =\sum_{n=1}^{\infty}
\frac{(u)_n}{(1)_n}\, z^n \, ,
\eea
 where $(a)_n= a(a+1)\cdots (a+n-1)= \frac{\Gamma(a+n)}{\Gamma(a)}$ is the Pochhammer symbol. Note that
$(1)_n= n!$. Comparing powers of $z^n$ gives
an expression for the generating function of $P(K\vert n)$ with respect $K$
\bea
\label{marg_K_alt.2}
\sum_{K=1}^{\infty} P(K\vert n)\, u^K= \frac{(u)_n}{(1)_n}= \frac{\Gamma(u+n)}{n!\, \Gamma(u)}\, .
\eea

\vspace*{0.5cm}
\noindent{\bf Marginal distributions of $M$.}
On the other hand, summing the joint distribution (\ref{JPDF_GF_3}) over $K = 1, 2, \cdots, + \infty$, one gets the generating function of the marginal distribution $P(M \vert n)$ of the maximum $M$ as
\bea \label{marg_M_1}
\sum_{n=1}^\infty P(M\vert n) \,z^n = \sum_{n=1}^\infty z^n \sum_{K=1}^\infty P(K,M\vert n) = \frac{z\,f(M)}{1-z\,F(M)} e^{- \ln(1-z F(M))} = \frac{z\,f(M)}{(1-z\,F(M))^2} \;.
\eea 
Expanding in powers of $z$ gives the well known result for extreme value statistics 
\bea \label{marg_M_2}
P(M\vert n) = n f(M) \left[ F(M)\right]^{n-1}= \frac{\partial}{\partial M}\left[ F(M)\right]^{n} \; .
\eea
This result has a simple interpretation: if the maximum $M$ is achieved by one of the $n$ entries, the rest of the $n-1$ entries must have their values less than or equal to $M$. 

To prove the result in Eq. (\ref{joint_factor}), we now compute the generating function of the RHS of Eq. (\ref{joint_factor}) using the results above derived for the marginal distributions. Using the result in (\ref{marg_M_2}), we get
\bea \label{GF_prod_1}
\sum_{n=1}^\infty P(K\vert n) P(M\vert n)\,z^n = \sum_{n=1}^\infty n f(M) \left[ F(M)\right]^{n-1} P(K \vert n)\, z^n = z f(M) \sum_{n=1}^\infty n \, P(K\vert n) \left[ z F(M)\right]^{n-1} \;.
\eea
To evaluate this sum, we differentiate the relation in Eq. (\ref{marg_K_2}) wrt $z$, which gives
\bea \label{marg_K_3}
\sum_{n=1}^\infty n\, P(K \vert n)\, z^{n-1} = \frac{1}{1-z} \, \frac{\left[ -\ln(1-z)\right]^{K-1}}{(K-1)!} \;.
\eea
Using (\ref{marg_K_3}) in (\ref{GF_prod_1}) with $z$ replaced by $z F(M)$ gives
\bea \label{GF_prod_2}
\sum_{n=1}^\infty P(K\vert n) P(M\vert n)\,z^n = \frac{z f(M)}{1-zF(M)}  \frac{\left[-\ln\left(1-z F(M)\right) \right]^{K-1}}{(K-1)!}  \;.
\eea
Subtracting (\ref{GF_prod_2}) from (\ref{JPDF_GF_3}) gives the identity
\bea \label{subtract}
\sum_{n=1}^\infty z^n \, \left[P(K,M\vert n) - P(K\vert n) P(M \vert n) \right] = 0 \;.
\eea
Since this relation is valid for arbitrary $z$, it follows that
\bea \label{joint_factor2}
P(K,M \vert n) = P(K \vert n) P(M \vert n) \quad, \quad {\rm for \; all} \quad n \geq 1 \;. 
\eea
Even though we have assumed here that the range of $X_i$'s is $[0, +\infty)$, it is clear that all the steps will go through even when $f(x)$ has a support over the full real line. Thus the two random variables $K$ and $M$ are completely uncorrelated for any IID sequence of arbitrary size. We find this result rather surprising as it could not have been easily guessed.

\section{Joint statistics of the number of records and the maximum in the dry-day model}\label{sec:IID_dry}

\begin{figure}[ht]
\includegraphics[width = 0.6\linewidth]{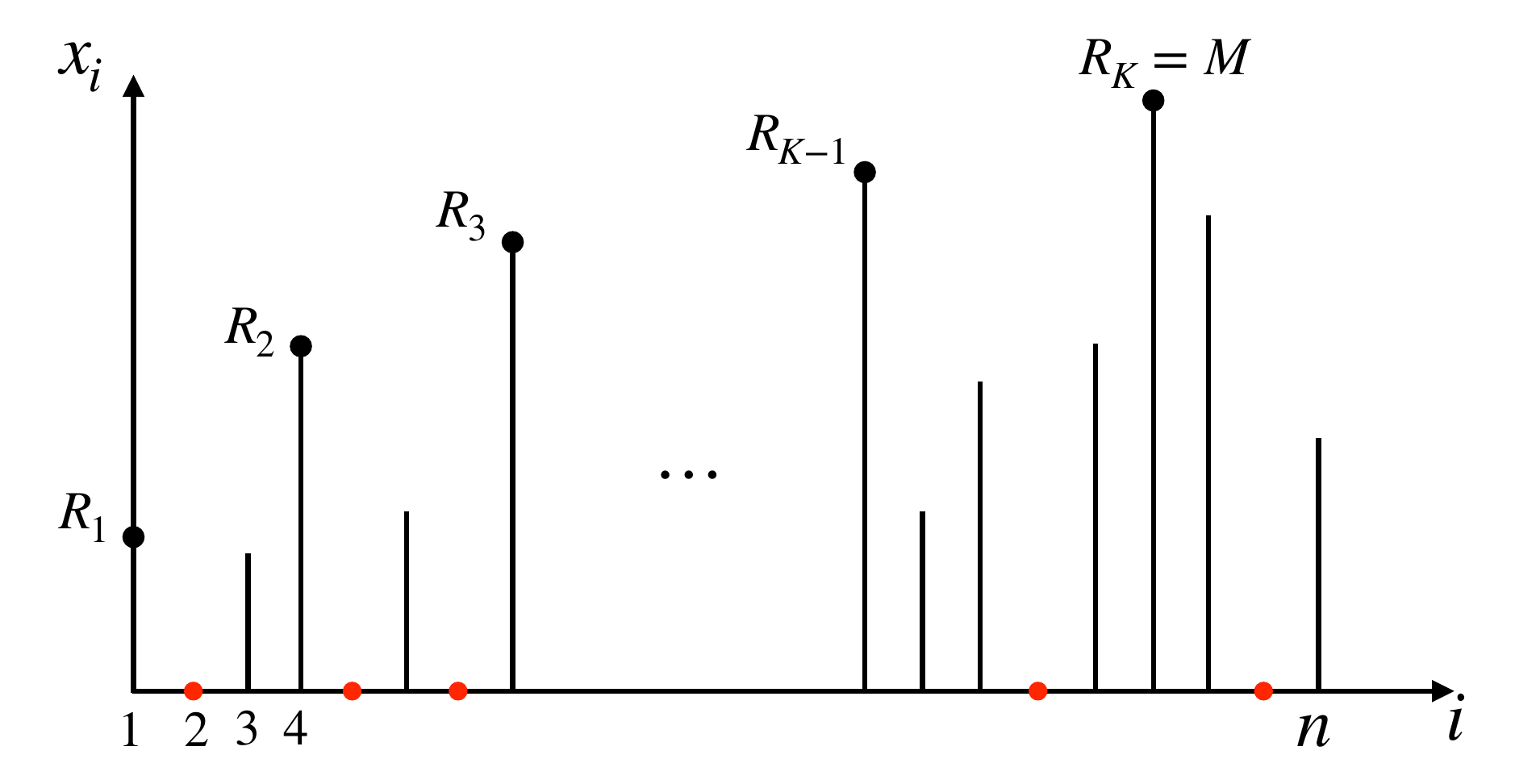}
\caption{A typical realization of an IID time series of $n$ entries and $K$ records in the dry-day model. Each entry is either zero (dry day), with probability $q$, indicated by a red dot, or drawn from a continuous distribution $f(x)$ (wet day), with probability $1-q$ [see Eq.~(1) in the Letter]. The record values $R_1,R_2,\ldots,R_K$ are indicated by black dots, with the last record $R_K=M$ corresponding to the global maximum.}\label{Fig_intro_dry}
\end{figure}

We now consider the same IID sequence of size $n$, but now some entries may be missing with probability $0\leq q \leq 1$. In the context of a precipitation time series, these missing entries may represent the days when there is no rainfall~\cite{Westcott77,BN92,DGP08,PK08,MBK19}. These are called ``dry days'' and they are somewhat similar to introducing dilute disorder on a $1d$ lattice (see Figure. \ref{Fig_intro_dry}). {We count records among wet-day observations only, and set $K=M=0$ when every day is dry.} For example, in Fig. \ref{Fig_intro_dry}, the red dots indicate the dry days and the record values are defined only for entries that represent ``wet days'' (i.e., not ``dry''). 

Our goal is to compute the connected correlation function of the number of records $K$ and the maximum $M$ in this disordered time series, for a fixed value of $q$. This connected correlation function is simply defined as
\bea \label{def_Cq}
C_q(n) = \langle K\, M\rangle(n) -\langle K \rangle(n) \langle M \rangle(n) 
\eea
For $q=0$, there is no dry day and we are back to the results of the previous section in Eq. (\ref{joint_factor2}), which says that $K$ and $M$ are completely uncorrelated, i.e., 
\bea \label{C_IID}
C_{q=0}(n) = 0 \quad, \quad {\rm for \, all} \quad n \geq 1 \;.
\eea 
For nonzero $q < 1$, we want to investigate whether nontrivial correlations get generated in the presence of a nonzero $q < 1$ and, if so, how $C_q(n)$ depends on the size $n$.

To proceed, we first define another random variable $n_W$ that represents the number of wet days in a given time series of size $n$. Clearly one has $0\leq n_W \leq n$. For the statistics of the number of records and the maximum, the relevant size of the sample is $n_W$. Hence we want to first compute the distribution $P(n_W \vert n)$ of $n_W$, for a given $n$. In the simplest ``Bernoulli'' model of records, the dry days occur independently at each day with a probability $0 \leq q \leq 1$ and, hence, $P(n_W \vert n)$ is simply given by the binomial distribution
\bea \label{Binomial}
P(n_W \vert n) = {n \choose n_W} (1-q)^{n_W}\, q^{n-n_W} \quad, \quad 0 \leq n_W \leq n \;.
\eea 
For $q=0$, $P(n_W \vert n) = \delta_{n_W,n}$ and, in this case, $K$ and $M$ are completely uncorrelated as shown in Eq. (\ref{C_IID}).

 For the simplicity of the computations, we will henceforth assume that the IID entries are each drawn from a distribution $f(x)$ that is supported over $x \in [0, + \infty)$. In the context of precipitation time series, this makes perfect sense since the rainfall is necessarily nonnegative. We will discuss	later what happens to $C_q(n)$ when one relaxes this constraint. 

To compute this correlation function $C_q(n)$ in Eq. (\ref{def_Cq}) we first compute the joint distribution $P_q(K,M \vert n)$. We first make an important observation. For $q=0$, as discussed in the previous section, we were using the convention that $n$ ranges over $1,2, \cdots$ and $K = 1, 2, \cdots$. In particular, the first entry, by definition, was a record. Thus, in this case, there is no possibility of having $K=0$. However, for $q>0$, there is a possibility that all entries are dry, which occurs with probability $q^n$. In this case there is no record, i.e., $K=0$. Since we are considering nonnegative variables, in this ``all dry'' configuration, the random variable $M$ can only take the value $M=0$. Consequently, for a generic $q$, we need to allow the possible range of $K$ to be $K=0,1,2, \cdots$ and set 
\bea \label{all_dry}
P_q(K=0,M \vert n) = q^n \delta(M) \;. 
\eea
The joint PDF $P_q(K,M \vert n)$ can be computed by first conditioning on $n_W$ and then summing over all possible $n_W$'s drawn from the binomial distribution (\ref{Binomial}). Hence we can write 
\bea \label{PqKM}
P_q(K,M \vert n) = \sum_{n_W = 0}^n P_c(K,M \vert n_W) {n \choose n_W} (1-q)^{n_W}\, q^{n-n_W} \;,
\eea  
where $P_c(K,M \vert n_W)$ is the conditional joint distribution of $K$ and $M$, given the number of nonzero entries $n_W$. Note that for $n_W>0$, we have
$P_c(K,M \vert n_W) = P(K \vert n_W) P(M \vert n_W)$ where $ P(K \vert n_W)$ and $P(M \vert n_W)$ are given in Eq. (\ref{stirling}) and (\ref{marg_M_2}) respectively with $n$ replaced by $n_W$. However, when $n_W = 0$, we have $P_c(K,M \vert n_W = 0)= \delta_{K,0} \delta(M)$, since in this case there is no nonzero entry and, hence, we have necessarily $K=0$ and also $M=0$ (as discussed above). Summarising, we have
 \bea \label{Pc}
 P_c(K,M \vert n_W) = 
 \begin{cases}
 & P(K \vert n_W) P(M \vert n_W) \quad {\rm for} \quad n_W>0 \;,\\
 & \\
 &  \delta_{K,0} \delta(M) \quad \quad \quad \quad  {\rm for} \quad n_W=0 \;.
 \end{cases}
\eea 
Substituting this result in Eq. (\ref{PqKM})  and separating out the $n_W = 0$ term from the sum gives
\bea \label{PqKM_2}
P_q(K,M \vert n) = \delta_{K,0}\delta(M)q^n + \sum_{n_W=1}^n P(K \vert n_W) P(M \vert n_W) {n \choose n_W} (1-q)^{n_W}\, q^{n-n_W} \;.
\eea
We now define the double generating function (GF) of $P_q(K,M \vert n)$ with respect to $K$ and $n$ as
\bea \label{def_GF1}
\sum_{n=0}^\infty v^n\,\sum_{K=0}^\infty u^K P_q(K,M \vert n) \;.
\eea 
Taking this double GF in Eq. (\ref{PqKM_2}) gives 
\begin{equation} \label{def_GF2}
\sum_{n=0}^\infty v^n\,\sum_{K=0}^\infty u^K P_q(K,M \vert n) = \frac{\delta(M)}{1-q\,v} + \sum_{n=0}^{\infty}\, v^n\sum_{K=0}^{\infty} u^K\, \sum_{n_W=1}^n P(K\vert n_W)\, P(M\vert n_W)\, {n\choose n_W}\, (1-q)^{n_W} q^{n-n_W} \;.
\end{equation}
In Section \ref{App_id} below we prove the identity
\begin{eqnarray}
&&\sum_{n=0}^{\infty}\, v^n\sum_{K=0}^{\infty} u^K\, \sum_{n_W=1}^n P(K\vert n_W)\, P(M\vert n_W)\, {n\choose n_W}\, (1-q)^{n_W} q^{n-n_W}
=\frac{\partial}{\partial M} \left[ \frac{(1-q v)^{u-1}}{(1- v g(M))^u}\right] \quad \label{A1.12} \\
&& \nonumber \\
&& {\rm where}\quad g(M)= q+(1-q) F(M) \quad, \quad {\rm with} \quad F(M) = \int_0^M f(x) \,dx \;,
\label{A1.11}
\end{eqnarray}
where $P(K\vert n)$ and $P(M\vert n)$ are the marginal distributions given respectively
in Eqs. (\ref{marg_K_alt.2}) and (\ref{marg_M_2}). Using this identity (\ref{A1.12}) in Eq. (\ref{def_GF2}), one finds
\bea \label{def_GF3}
\sum_{n=0}^\infty v^n\,\sum_{K=0}^\infty u^K P_q(K,M \vert n) = \frac{\delta(M)}{1-q\,v} + \frac{\partial}{\partial M} \left[ \frac{(1-q v)^{u-1}}{(1- v g(M))^u}\right] \;. 
\eea
From this expression (\ref{def_GF3}), we can also compute the GFs of the marginal distributions $P_q(K\vert n) = \int_0^\infty \,P_q(K,M \vert n) dM$ and $P_q(M \vert n) = \sum_{K \geq 0} P_q(K,M \vert n)$. Integrating (\ref{def_GF3}) with respect to $M$ and using the fact that $F(M \to \infty) = 1$ and $F(M \to 0) = 0$, we get
\bea \label{marginPqK}
\sum_{K,n = 0}^\infty P_q(K \vert n) u^K\, v^n =\frac{(1-qv)^{u-1}}{(1-v)^u} \;.
\eea
Also setting $u=1$ in Eq. (\ref{def_GF3}) gives
\bea \label{marginPqM}
\sum_{n=0}^\infty P_q(M \vert n) \,v^n = \frac{\delta(M)}{1-q\,v} + \frac{\partial}{\partial M} \left[ \frac{1}{1-v\,g(M)}\right] \;,
\eea
where $g(M)$ is defined in Eq. (\ref{A1.11}). Expanding in powers of $v$, we get 
\bea \label{PqMn}
P_q(M \vert n) = \delta(M)\,q^n + \frac{\partial}{\partial M} \left[ g^n(M)\right] \;.
\eea

Armed with these results for the joint distribution of $K$ and $M$ in Eq. (\ref{def_GF3}) and their respective marginal distributions in (\ref{marginPqK}) and (\ref{PqMn}), our next goal is to compute the generating function of the connected correlation function in Eq. (\ref{def_Cq})
 \bea \label{GF_Cq}
\tilde C_q(v) = \sum_{n=0}^\infty C_q(n)\,v^n = \sum_{n=0}^\infty \left[ \langle K\, M\rangle(n) -\langle K \rangle(n) \langle M \rangle(n) \right] v^n \;.
\eea 
To compute the RHS of Eq. (\ref{GF_Cq}) we need to compute the quantity $P_q(K,M \vert n) - P_q(K\vert n) P_q(M \vert n)$. Hence our next step is to compute the double GF  
\bea \label{GF_conn}
\sum_{K,n = 0}^\infty \left[P_q(K,M \vert n) - P_q(K\vert n) P_q(M \vert n)\right] u^K\,v^n 
\eea
The first term is already given in Eq. \eqref{def_GF3}. We now compute the second term as follows. Using Eq. (\ref{PqMn}) we express the second term as follows
\bea \label{sec_1}
\sum_{K,n = 0}^\infty P_q(K\vert n) P_q(M \vert n)\, u^K\,v^n &=& \sum_{K,n = 0}^\infty P_q(K\vert n) \left[ \delta(M)\,q^n + \frac{\partial}{\partial M} \left[ g^n(M)\right]\right] \, u^K\,v^n \\
&=& \delta(M) \sum_{K,n=0}^\infty P_q(K\vert n) u^K (q\,v)^n + \frac{\partial}{\partial M} \left[\sum_{K,n=0}^\infty P_q(K \vert n) u^K \,(v\, g(M))^n \right] \;.
\eea
Using now Eq. (\ref{marginPqK}) we can evaluate the sums over $K$ and $n$ explicitly, which leads to
\bea \label{sec_2}
\sum_{K,n = 0}^\infty P_q(K\vert n) P_q(M \vert n)\, u^K\,v^n = \delta(M) \frac{(1-q^2\,v)^{u-1}}{(1-q\,v)^u} + \frac{\partial}{\partial M} \left[ \frac{(1-q\,v\,g(M))^{u-1}}{(1-v\,g(M))^{u}}\right]
\eea 
Subtracting Eq. (\ref{sec_2}) from Eq. (\ref{def_GF3}) then gives
\bea \label{GF_conn2}
\sum_{K,n = 0}^\infty \left[P_q(K,M \vert n) - P_q(K\vert n) P_q(M \vert n)\right] u^K\,v^n  &=& \delta(M) \left[\frac{(1-q\,v)^{u-1}-(1-q^2\,v)^{u-1}}{(1-q\,v)^u}\right] \nonumber \\
&+& \frac{\partial}{\partial M} \left[ \frac{(1-q\,v)^{u-1} - (1-q\,v\,g(M))^{u-1}}{(1-v\,g(M))^u}\right] \;.
\eea
Note that when $q \to 0$, the RHS vanishes, thus recovering the result in Eq. (\ref{joint_factor2}) that, in this case, $K$ and $M$ are completely uncorrelated. 

We are now ready to compute the RHS of Eq. (\ref{GF_Cq}). Multiplying Eq. (\ref{GF_conn2}) by $M$ and integrating over $M \in [0, +\infty)$, via integration by parts, gives
\bea \label{intM_1}
&&\sum_{K,n=0}^\infty \left[\int_0^\infty M P_q(K,M \vert n) \, dM - P_q(K\vert n) \int_0^\infty M\,P_q(M\vert n)\,dM\right]\, u^K\, v^n \\
&&= \int_0^\infty dM \left[ \frac{1}{1-q\,v\,g(M)} \left[ \frac{1-q\,v\,g(M)}{1-v\,g(M)}\right]^u - \frac{1}{1-q\,v} \left[ \frac{1-q\,v}{1-v\,g(M)}\right]^u  \right] \;.
\eea
We now take a derivative with respect to $u$ on both sides and set $u=1$. We use the following identity
\bea \label{id_u}
\lim_{u \to 1} \frac{d}{du} x^u = x\, \ln x \;.
\eea
This then gives a rather compact exact result, valid for general $f(x)$ with positive support (and with a finite first moment)
\begin{equation} \label{GF_Cq_final}
\tilde C_q(v) = \sum_{n=0}^\infty C_q(n)\,v^n = \sum_{n=0}^\infty \left[ \langle K\, M\rangle(n) -\langle K \rangle(n) \langle M \rangle(n) \right] v^n  \\
=\int_0^\infty \frac{dM}{1-v\,g(M)} \ln \left[ \frac{1-q \, v \, g(M)}{1-q\,v}\right] \;,
\end{equation}
where we recall that
\bea \label{rec_g}
g(M) = q + (1-q)\,F(M) = 1 - (1-q)\bar{F}(M) \quad {\rm where} \quad \bar{F}(M) = 1 - F(M) = \int_M^\infty f(x)\,dx \;.
\eea
This formula in Eq. (\ref{GF_Cq_final}) is the main result of this section, which is valid for any continuous parent distribution $f(x)$ with a positive support. 

In the Sec.~\ref{subsec:large_n}, we will use this formula to analyse the small and large $n$ behavior of $C_q(n)$. However, before we proceed, we give a proof of the identity in Eq.~\eqref{A1.12} below.

\section{Proof of the identity in Eq. (\ref{A1.12})}\label{App_id}

In this section, we prove the following identity [stated earlier in Eq.~\eqref{A1.12}]
\begin{eqnarray}
&&\sum_{n=0}^{\infty}\, v^n\sum_{K=0}^{\infty} u^K\, \sum_{n_W=1}^n P(K\vert n_W)\, P(M\vert n_W)\, {n\choose n_W}\, (1-q)^{n_W} q^{n-n_W}
=\frac{\partial}{\partial M} \left[ \frac{(1-q v)^{u-1}}{(1- v g(M))^u}\right] \quad \label{A1.0}\\
&& \nonumber \\
&& {\rm where}\quad g(M)= q+(1-q) F(M) \quad, \quad {\rm with} \quad F(M) = \int_0^M f(x) \,dx \;,
\label{A1.1}
\end{eqnarray}
where $P(K\vert n)$ and $P(M\vert n)$ are the marginal distributions given respectively
in Eqs. (\ref{marg_K_alt.2}) and (\ref{marg_M_2}). 

We start with the left hand side (LHS) of Eq. (\ref{A1.0}). Noting that $P(K=0\vert n_W)=0$ for any $n_W\ge 1$, we note
that the sum over $K$ runs over $K=1,2\cdots$. Using the result from Eq. (\ref{marg_K_alt.2}) and the explicit form of
$P(M\vert n_W)$ from Eq. (\ref{marg_M_2}), we can express the LHS of
Eq. (\ref{A1.1}) as
\bea
\label{A1.2}
{\rm LHS} &= & \frac{\partial}{\partial M} \left[ \sum_{n=0}^{\infty} (qv)^n\, \sum_{n_W=1}^n \frac{(u)_{n_W}}{(1)_{n_W}}\,
{n\choose n_W} \left[ \frac{(1-q) F(M)}{q}\right]^{n_W}\right]\, \nonumber \\
&=& \frac{\partial}{\partial M} \left[ \sum_{n=0}^{\infty} (qv)^n\, \sum_{n_W=0}^n \frac{(u)_{n_W}}{(1)_{n_W}}\,
{n\choose n_W} \left[ \frac{(1-q) F(M)}{q}\right]^{n_W}\right]\, ,
\eea 
where in the second line we have inserted the $n_W=0$ term for algebraic convenience. Note that we are allowed to do 
this since the term corresponding to $n_W=0$
inside the parenthesis is simply $\sum_{n=0}^{\infty} (qv)^n= 1/(1-qv)$ which is independent of $M$ and hence its derivative with
respect to $M$ just gives zero. To perform the double sum over $n$ and $n_W$ in Eq. (\ref{A1.2}), we interchange the sums, namely
we run $n_W$ from $0$ to $\infty$, while $n$ runs from $n_W$ to $\infty$. In other words
\bea 
\label{A1_inter}
{\rm LHS}= \frac{\partial}{\partial M} \left[\sum_{n_W=0}^{\infty} \frac{(u)_{n_W}}{(1)_{n_W}}\left[ \frac{(1-q) F(M)}{q}\right]^{n_W}\, \sum_{n=n_W}^{\infty}
{n\choose n_W} (qv)^n \right]\, .
\eea
Now the sum over $n$ can be performed trivially using the identity
\bea
\sum_{n=n_W}^{\infty} {n \choose n_W}\, x^n = \frac{x^{n_W}}{(1-x)^{n_W+1}}\, .
\label{A1_ident.1}
\eea 
Substituting this identity in (\ref{A1_inter}) gives
\bea
\label{A1.3}
{\rm LHS}= \frac{1}{1-q v}\, \frac{\partial}{\partial M}\left[\sum_{n_W=0}^{\infty} \frac{(u)_{n_W}}{(1)_{n_W}}\, 
\left(\frac{v(1-q) F(M)}{1-qv}\right)^{n_W}\right]\, .
\eea
The sum over $n_W$ can now be performed explicitly and we get
\bea
{\rm LHS}=\frac{1}{1-q v}\, \frac{\partial}{\partial M}\left[\frac{1}{\left[1 - \frac{v(1-q) F(M)}{1-qv}\right]^u}\right]\, .
\label{A1.4}
\eea
The RHS of Eq. (\ref{A1.4}) coincides with the RHS of (\ref{A1.0}), thus proving the identity.

\section{Asymptotic behaviours of $C_q(n)$ for large $n$}
\label{subsec:large_n}
In this section, we start from Eq. (\ref{GF_Cq_final}) and derive the behavior of $C_q(n)$ for large $n$.
We will show  below that, for large $n$, the correlator $C_q(n)$ decays as a power law, with an exponent that depends on the tails of the parent distribution $f(x)$.

To extract the large $n$ behavior of $C_q(n)$ from the exact generating function in Eq. (\ref{GF_Cq_final}), we need to investigate the limit $v \to 1$.
We first consider the case when the parent distribution $f(x)$ is supported over the full positive real line, i.e., $x \in [0,+\infty)$. It is convenient to first make the substitution $v=1-s$ on both sides of Eq. (\ref{GF_Cq_final}). The integrand on the RHS then becomes
\bea \label{integrand}
I(M,s) =\frac{1}{1-(1-s)\,g(M)} \ln \left[ \frac{1-q \, (1-s) \, g(M)}{1-q\,(1-s)}\right] \;.
\eea
Using $g(M) = 1-(1-q)\bar{F}(M)$, this integrand can be expressed as
\bea \label{integrand_2}
I(M,s) = \frac{1}{s + (1-q)\bar{F}(M) - (1-q)s\,\bar{F}(M) } \ln \left(1 + \frac{q(1-q)(1-s)\bar{F}(M)}{1-q(1-s)} \right) \;.
\eea
We first note that, exactly at $s=0$, the integrand reduces to 
\bea  \label{integrand_3}
I(M,0) = \frac{1}{(1-q)\bar{F}(M)} \ln \left(1 + q \bar{F}(M) \right) \;.
\eea
Then clearly the integral $\int_0^\infty I(M,0)\,dM$ diverges since $\bar{F}(M) = \int_M^\infty f(x)\,dx \to 0$ as $M \to \infty$. This shows that $\sum_{n \geq 0} C_q(n)$ is divergent. Hence, to extract the large $n$ behaviour of $C_q(n)$, we need to investigate the limit $s \to 0$, which we expect to be singular. 
From the expression of the integrand in (\ref{integrand_2}), we see that, as $s \to 0$, the dominant contribution to the integral emerges from the large $M$ limit of $I(M,s)$ where $\bar{F}(M) \to 0$. Hence, we need to analyse the scaling regime of $I(M,s)$ when $s \to 0$ and $\bar{F}(M) \to 0$ simultaneously, but keeping $s/\bar{F}(M)$ fixed. In this scaling limit, the integrand $I(M,s)$ reduces to
\bea \label{integrand_4}
I(M,s) \approx \frac{q\,\bar{F}(M)}{s+(1-q)\bar{F}(M)} \;.
\eea
This approximation holds of course only for large $M$. Therefore, the integral on the RHS of Eq. (\ref{GF_Cq_final}) can be split into a regular and a singular part by writing 
\bea \label{int_Cq}
\sum_{n=0}^\infty C_q(n)(1-s)^n \approx \int_0^\Lambda I(M,s)\, dM + \int_\Lambda^\infty \frac{q\,\bar{F}(M)}{s+(1-q)\bar{F}(M)} \, dM \;,
\eea
where we have introduced a cut-off $\Lambda \gg 1$, such that $I(M,s)$ can be approximated by the RHS of (\ref{integrand_4}) for $M > \Lambda$. To extract the leading large $n$ behaviour of $C_q(n)$, we only need to analyse the singular term [the second term on the RHS of Eq. (\ref{int_Cq})]. Neglecting the regular term we then arrive at the approximation
\bea \label{Cq_largen_1}
\sum_{n=0}^\infty C_q(n)\,(1-s)^n \approx  \int_{0}^\infty C_q(n)\, e^{-s\,n} \, dn \approx q\,\int_\Lambda^\infty dM \frac{\bar{F}(M)}{s+(1-q) \bar{F}(M)} \;.
\eea
where we replaced the sum in the first term by an integral, valid in the small $s$ limit. This then converts effectively the generating function into a Laplace transform. Inverting the Laplace transform explicitly gives
 \bea \label{Cq_largen_3}
C_q(n) \approx q \int_\Lambda^\infty dM \, \bar{F}(M)\,e^{-n(1-q)\, \bar{F}(M)} =    q \int_0^\infty dM \, \bar{F}(M)\,e^{-n(1-q)\, \bar{F}(M)} - q \int_0^\Lambda dM \, \bar{F}(M)\,e^{-n(1-q)\, \bar{F}(M)}\;.
\eea
In the large $n$ limit the second term becomes subleading, as it decays at least exponentially. Hence the leading singular behavior of $C_q(n)$ is  
given by the compact formula
\bea \label{Cq_largen_4}
C_q(n) \approx q \int_0^\infty dM \, \bar{F}(M)\,e^{-n(1-q)\, \bar{F}(M)} \;.
\eea
For parent distributions with a finite support $x \in [0,x^*]$ one can follow an exactly similar analysis and the large $n$ behavior is given by an analogous formula but with the upper limit of integral in Eq. (\ref{Cq_largen_4}) replaced by $x^*$, i.e.,
\bea \label{Cq_largen_5}
C_q(n) \approx q \int_0^{x^*} dM \, \bar{F}(M)\,e^{-n(1-q)\, \bar{F}(M)} \;.
\eea
We analyse below this formula for large $n$ for different choices of parent distributions $f(x)$ and we will see that, typically, the correlator $C_q(n)$
decays as a power law, with an exponent that depends on the behavior of $f(x)$ near $x = x^*$.

\vspace*{0.5cm}
\noindent{\bf Purely exponential distribution $f(x) = e^{-x}$, for $x>0$.} In this case the integral in Eq. (\ref{Cq_largen_4}) can be performed explicitly, giving the leading order behavior for large $n$
\bea \label{Cq_Exp}
C_q(n) \approx \frac{q}{(1-q)n} \;.
\eea
Thus, the covariance between $K$ and $M$ decays as $1/n$. We also remark that when $q \to 0$, the correlator vanishes, as expected from Eq. (\ref{C_IID}). 

\vspace*{0.5cm}
\noindent{\bf Distribution with a stretched exponential tail $f(x) \approx A\,e^{-x^\gamma}$, as $x \to \infty$, with $\gamma>0$.} In this case, for large $M$, 
one has the leading asymptotic behavior
\bea \label{barF_stret}
\bar{F}(M) = \int_M^\infty f(x)\,dx \approx \frac{A}{\gamma}\, M^{1-\gamma} \,e^{-M^\gamma} \;.
\eea
The dominant contribution in Eq. (\ref{Cq_largen_4}) for large $n$ comes from the regime where $\bar{F}(M)\to 0$. Hence we can substitute the leading large $M$ behaviour of $\bar{F}(M)$ from Eq. (\ref{barF_stret}) in the integral in Eq. (\ref{Cq_largen_4}). This leads to
\bea \label{Cq_stret}
C_q(n) \approx \frac{q}{\gamma\,(1-q)\,n} \, \frac{1}{\left[ \ln((1-q)n)\right]^{1- \frac{1}{\gamma}}} \;.
\eea
Therefore the correlator $C_q(n)$ decays as $1/n$ (with logarithmic corrections that depend on $\gamma$). For $\gamma  = 1$, it reduces to the result in (\ref{Cq_Exp}). Also for $q=0$ one recovers $C_{q=0}(n) = 0$, as expected from Eq. (\ref{C_IID}). Interestingly the leading large $n$ behavior of $C_q(n)$ in Eq. (\ref{Cq_stret}) does not depend on the amplitude $A$ in the parent distribution $f(x) \approx A \,e^{-x^\gamma}$.  

\vspace*{0.5cm}
\noindent{\bf Distribution with a power law tail $f(x) \approx B/x^{\mu +1}$ as $x \to \infty$ with $\mu >0$.} In this case, 
\bea \label{barF_power}
\bar{F}(M) = \int_M^\infty f(x)\,dx \approx \frac{B}{\mu}\, \frac{1}{M^{\mu}} \;.
\eea
Substituting this behavior in Eq. (\ref{Cq_largen_4}), we get
\bea \label{Cq_power}
C_q(n) \approx \frac{q\, B}{\mu}\, \int_0^\infty \frac{dM}{M^{\mu}} \, e^{-(1-q)\,n \frac{B}{\mu\, M^\mu}} \;.
\eea
This integral diverges for $0<\mu < 1$ due to the fact that the integrand behaves as $\sim M^{-\mu}$ for large $M$. Physically this makes sense because 
for $0<\mu \leq 1$ the correlator does not exist since, typically, the random variable $M$ denoting the value of the global maximum itself has a power law distribution $P(M)\sim 1/M^{\mu +1}$ {indicating that all the moments of $M$, including its average, diverge.} As a result, the correlation function $C_q(n)$ also does not exist for any $n$. Hence we restrict ourselves to the case $\mu > 1$. In this case the integral in Eq. (\ref{Cq_power}) is convergent and we get
\bea \label{Cq_power_large}
C_q(n) \approx \frac{\tilde B}{((1-q)n)^{1-\frac{1}{\mu}}} \quad, \quad {\rm where} \quad \tilde B = \frac{q\, B^{1/\mu}}{\mu^{1+\frac{1}{\mu}}} \Gamma\left( 1 - \frac{1}{\mu}\right) \;.
\eea 
Note that the decay exponent $1-1/\mu \to 1$ as $\mu \to \infty$, which is consistent with the result for the exponential case in Eq. (\ref{Cq_Exp}).

\vspace*{0.5cm}
\noindent{\bf Distribution with a bounded support over $[0,x^*]$.} Here we consider a parent distribution that is supported over $[0,x^*]$ and for convenience we will choose $x^*=1$ without any loss of generality. We assume that, as $x \to 1$, the distribution behaves as $f(x) \approx C\,(1-x)^\alpha$ with $\alpha > - 1$ (so that it is integrable around $x = 1$). In this case 
\bea \label{barF_finite}
\bar{F}(M) = \int_M^1 f(x)\,dx \approx \frac{C}{\alpha + 1}(1-M)^{\alpha + 1} \quad, \quad M \to 1 \;.
\eea
Inserting this behaviour in Eq. (\ref{Cq_largen_5}) and using the fact that the integral for large $n$ is dominated by the vicinity of $M=1$  one finds 
\bea \label{barF_finite_large}
C_q(n) \approx \frac{q \,C}{\alpha+1} \int_0^{1} dM \, (1-M)^{\alpha + 1}\,e^{-n(1-q) \frac{C}{\alpha+1}\, (1-M)^{\alpha  +1 }} \;.
\eea
Evaluating this integral in the large $n$ limit gives

\bea\label{barF_finite_large2}
C_q(n) \approx \frac{\tilde C}{((1-q)n)^\frac{\alpha+2}{\alpha+1}} \quad, \quad \tilde C = \frac{q\,\Gamma\left( \frac{1}{\alpha+1}\right)}{C^{\frac{1}{\alpha+1}}(\alpha+1)^{\frac{2\alpha+1}{\alpha+1}} } \;.
\eea
For example, for the case of a uniform distribution $f(x) = 1$ for $0 \leq x \leq 1$, which corresponds to $\alpha = 0$ and $C=1$, we get 
\bea \label{Cq_unif}
C_q(n) \approx \frac{q}{[(1-q)n]^2} \;.
\eea

\section{The random record model: the continuum limit} \label{sec:continuum}

In the limit when $q\to 1$, there are very few wet days in a sequence of $n$ days. In fact, the mean number of wet days is just $(1-q)\,n$. Therefore a natural limit is to take $q \to 1$ and $n \to \infty$, while keeping the product $(1-q)\,n = t$ fixed. In this limit, the rare wet days form a Poisson point process with unit rate. In the literature on record statistics, this limiting process is known as ``the random record model''~\cite{Westcott77,BN92,DGP08,PK08}. It is natural to ask if our correlation function $C_q(n)$ also converges to a continuous function $C(t)$ in this limit. Indeed, we will show that in the scaling limit the correlator takes the scaling form
\bea \label{Cq_scaling}
C_q(n) \approx C(t = (1-q)\,n) \;,
\eea 
where we show that the Laplace transform of $C(t)$ is given exactly by
\bea \label{Laplace_C}
\int_0^\infty C(t)\, e^{-st} \, dt = \int_0^\infty \frac{dM}{s + \bar{F}(M)} \ln \left(1+ \frac{\bar F(M)}{s+1} \right) \;,
\eea
where we recall once more that $\bar{F}(M) = \int_M^\infty f(x)\,dx$. Note that this result holds for a parent distribution $f(x)$ that is supported over the full positive line. For $f(x)$ with a finite support $[0,x^*]$, it is easy to see that the same result as in Eq. (\ref{Laplace_C}) holds with the upper limit $+\infty$ replaced by $x^*$, i.e.,
\bea \label{Laplace_C_finite}
\int_0^\infty C(t)\, e^{-st} \, dt = \int_0^{x^*} \frac{dM}{s + \bar{F}(M)} \ln \left(1+ \frac{\bar F(M)}{s+1} \right) \;.
\eea

To establish this result, we again start from our main exact result in Eq. (\ref{GF_Cq_final}). We set $p=1-q$. Hence in this continuum limit as $q\to 1$, the fraction of wet days $p \to 0$. We set $v = 1-p s$ on both the LHS and the RHS of Eq.~(\ref{GF_Cq_final}). On the RHS, we see that in this limit $p\to 0$ with $s$ fixed, the leading behaviour of the integral is given by
\bea \label{RHS_sto0}
{\rm RHS} \sim \frac{1}{p}\int_0^\infty \frac{dM}{s + \bar{F}(M)} \ln \left(1+ \frac{\bar F(M)}{s+1} \right) \;.
\eea
On the other hand, the LHS behaves as
\bea \label{LHS_sto0}
\sum_{n=1}^\infty C_q(n)\, v^n = \sum_{n=1}^\infty C_q(n) (1-p\,s)^n \approx \sum_{n=1}^\infty C_q(n)\,e^{-sn p} \approx \frac{1}{p}\int_0^\infty C(t)\,e^{-st}\,dt \;,
\eea
where we replaced the summation, in the third term, by an integral over $t= n\,p$, which is allowed in the $p \to 0$ limit. Comparing the leading behaviours of the LHS and RHS as $p \to 0$ in Eqs. (\ref{LHS_sto0}) and (\ref{RHS_sto0}), we get the result in Eq. (\ref{Laplace_C}).  

\begin{figure}[t]
\includegraphics[width=0.5\linewidth]{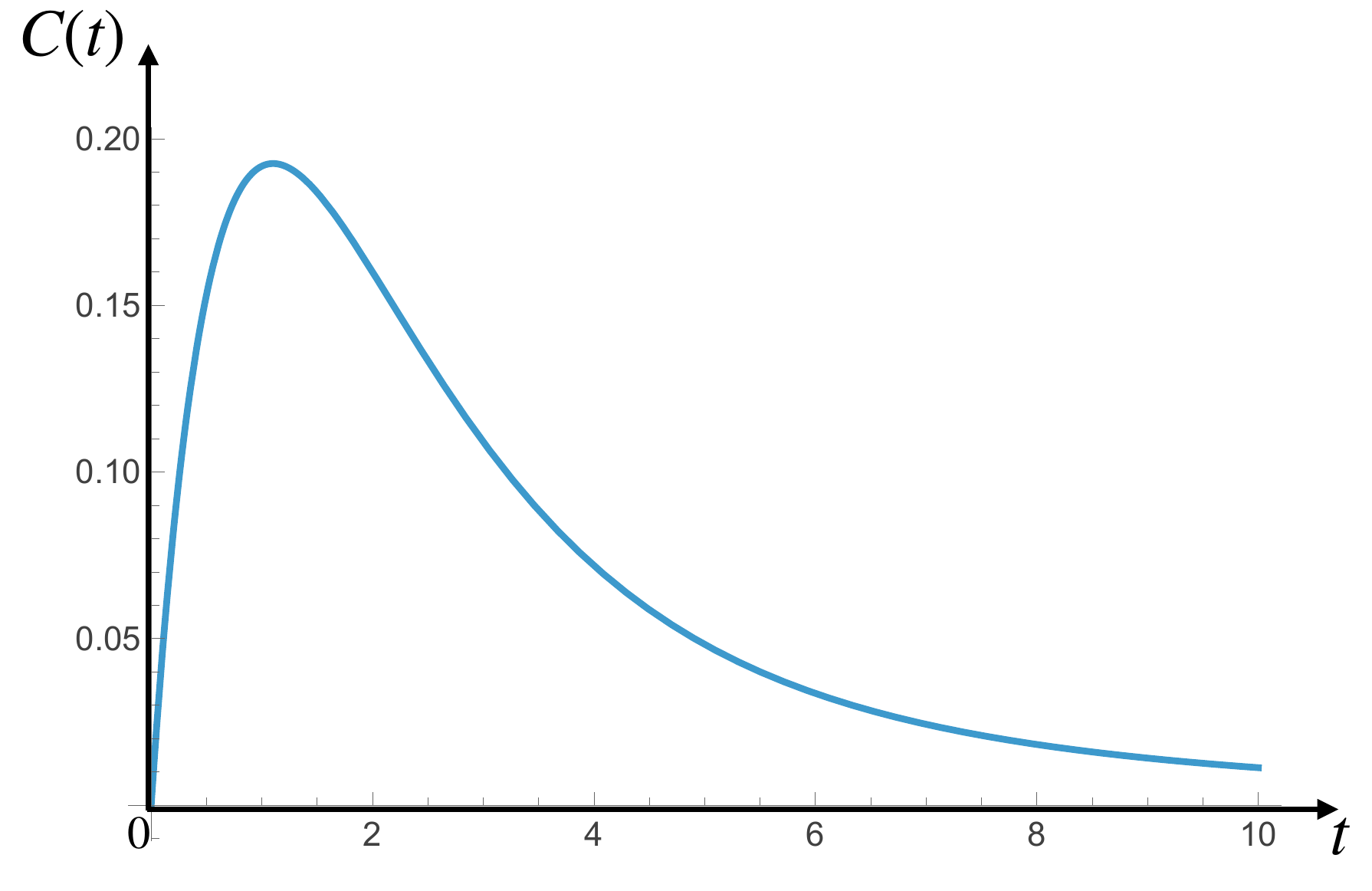}
\caption{A plot of $C(t)$ vs $t$ given in Eq. (\ref{C_cont_unif4}).}\label{Fig_Ct_exp}
\end{figure}

For a general parent distribution $f(x)$, the Laplace transform can be formally inverted, leading to Eq. (EM.14) in the End Matter. Here we write this expression in a slightly different form, namely   
\bea \label{Ct_SM}
C(t)
=
\int_0^\infty dM\,
e^{-t\bar F(M)} \int_{F(M)t}^t \frac{(1-e^{-u})}{u}\,{du}\;,
\eea
where we recall that $F(M) = 1-{\bar F}(M)= \int_0^M f(x)\,dx$. Note that it is generally difficult to perform the remaining integral over $M$ in (EM.14) to obtain a closed form expression for $C(t)$. A simpler case is the uniform distribution $f(x) = 1$ for $0 \leq x \leq 1$. In this case $F(M)=M$ leading to
\bea \label{Ct_SM2}
C(t) = e^{-t}\int_0^1 dM e^{M\,t} \int_{M\,t}^t \frac{(1-e^{-u})}{u}\,{du} \;.
\eea
Making a change of variable $v = M\,t$ and performing the integral (over $v$) by parts, we get the expression given in Eq. (EM17) in the End Matter, namely,
\bea \label{C_cont_unif4}
C(t) = \frac{2}{t}\,e^{-t}\int_0^t \frac{(\cosh(v) - 1)}{v}\,dv \quad, \quad t \geq 0 \;.
\eea
In Fig. \ref{Fig_Ct_exp} we show a plot of this function. It has the following asymptotic behaviours
\bea \label{Ct_asympt}
C(t) \approx
\begin{cases}
&\dfrac{t}{2} - \dfrac{t^2}{2} + O(t^3) \quad, \quad t \to 0 \;, \\
& \\
& \dfrac{1}{t^2} + O(t^{-3} )\quad, \quad \quad\, t \to +\infty \;.
\end{cases}
\eea
The power law decay $1/t^2$ for large $t$ matches smoothly with the asymptotic large $n$ result in Eq. (\ref{Cq_unif}) in the limit $q \to 1$, $n \to \infty$ with $n(1-q) = t \gg 1$ fixed.  

For a generic $f(x)$, while $C(t)$ can be expressed formally as in Eq. (\ref{Ct_SM}), the small and large $t$ of $C(t)$ are conveniently extracted from the Laplace transform in Eq. (\ref{Laplace_C}) explicitly, as we show now.

\vspace*{0.5cm}
\noindent{\bf Small $t$ asymptotics.} To extract the small $t$ behaviour of $C(t)$, we need to investigate the large $s$ behavior of the RHS of Eq. (\ref{Laplace_C}). In this limit, the leading large $s$ behavior of the RHS is given by
\bea \label{large_s}
{\rm RHS} \to \frac{\mu_1}{s^2} \quad, \quad {\rm where} \quad \mu_1 = \int_0^\infty \bar{F}(M) dM \;.
\eea
Note again that, for a distribution $f(x)$ with a finite support over $[0,x^*]$, the upper limit of the integral in $\mu_1$ needs to be replaced by $x^*$. Inverting trivially the Laplace transform we then get a universal linear growth for small $t$ of the correlation function $C(t)$, namely
\bea \label{Ct_smallt}
C(t) \approx \mu_1 \, t \quad, \quad t \to 0 \;.
\eea
Replacing $t = n(1-q)$ we see that $C_q(n) \approx \mu_1 n (1-q)$ for $n(1-q) \ll 1$. This matches smoothly with the results for $C_q(1)$ and $C_q(2)$ in Eqs. (EM.4) and (EM.6) (in the End Matter) in the limit $q \to 1$.  

\vspace*{0.5cm}
\noindent{\bf Large $t$ asymptotics.} To extract the large $t$ behavior of $C(t)$, we need to investigate the small $s$ limit of the RHS in Eq. (\ref{Laplace_C}). As argued previously for $q<1$ case in Section \ref{subsec:large_n}, the leading small $s$ behavior of the RHS is captured in the scaling limit $s \to 0$ and $\bar F(M) \to 0$ with their ratio fixed. Following the same line of analysis as in Section \ref{subsec:large_n}, we then get
\bea \label{Ct_larget}
\int_0^\infty C(t)\,e^{-st}\, dt \approx \int_0^\infty dM\, \frac{\bar F(M)}{s+\bar F(M)} \;.
\eea
Inverting this Laplace transform gives the leading large $t$ behaviour of $C(t)$ as
\bea \label{Ct_larget_2}
C(t) \approx \int_0^\infty dM\, \bar F(M)\, e^{-\bar F(M)\,t}  \;.
\eea
Note that the formula for $C_q(n)$ for large $n$ and $q$ fixed in Eq. (\ref{Cq_largen_4}) coincides with the expression of $C(t)$ in (\ref{Ct_larget_2}) in the limit when $q \to 1$, $n \to \infty$ with $t=n(1-q) \gg 1$. One can then analyse this integral in Eq. (\ref{Ct_larget_2}) for large $t$ exactly as we did in Section \ref{subsec:large_n}. We can in fact read off the large $t$ asymptotics from the results in Section \ref{subsec:large_n} by replacing the factor $(1-q)n$ by $t$ and taking the limit $q \to 1$. Summarising, we find that for any parent distribution $f(x)$, the correlator $C(t)$ typically decays as a power law (with additional logarithmic corrections in special cases), namely
\bea \label{summary_Ct}
C(t) \sim \frac{1}{t^\theta} \quad, \quad {\rm as} \quad t \to \infty \;,
\eea
where the exponent $\theta$ depends on the tail of the parent distribution $f(x)$. {Skipping further details, let us summarise the main results:} 
\begin{itemize}
\item[(i)] {Exponential/stretched exponential distribution $f(x) \sim e^{-x^\gamma}$: in this case we get $\theta = 1$ with logarithmic corrections for $\gamma \neq 1$ (see for example Eq. (\ref{Cq_stret})). In other words 
\bea \label{Ct_exp}
C(t) \approx \frac{1}{\gamma} \frac{1}{t\, (\ln t)^{1-1/\gamma}} \;.
\eea
}
\item[(ii)]{Power law distribution $f(x) \sim 1/x^{1+\mu}$ with $\mu >0$: in this case $\theta = 1 - \frac{1}{\mu}$ for $\mu > 1$. As before, for $0<\mu \leq 1$ the correlation function $C(t)$ does not exist. 
} 
\item[(iii)]{Bounded distribution $f(x) \sim (1-x)^\alpha$ with $\alpha>-1$. In this case $\theta = \frac{\alpha+2}{\alpha+1}$.}

\end{itemize}

\section{Analysis of precipitation data}

As mentioned in the main text, we analyze daily precipitation records from
the Climate Data Center of the German Weather Service, \emph{Deutscher
Wetterdienst} (DWD, \url{https://www.dwd.de}). The dataset contains records
from 6,105 weather stations, of which 573 provide complete daily coverage
over the 40-year period 1974--2013. To limit inter-station variability in
the fraction of dry days (days with zero precipitation), we further restrict
the analysis to stations for which this fraction lies in the interval
$[0.48, 0.52]$. This leaves 232 stations. For the combined dataset, the
fraction of dry days is $q = 0.501569\dots$.

\begin{figure}
    \centering
    \includegraphics[width=0.48\textwidth]{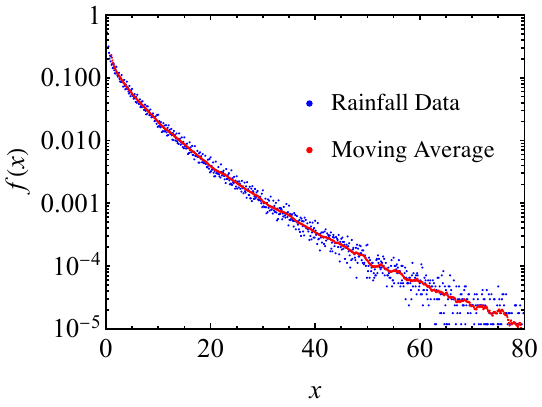}~~\includegraphics[width=0.46\textwidth]{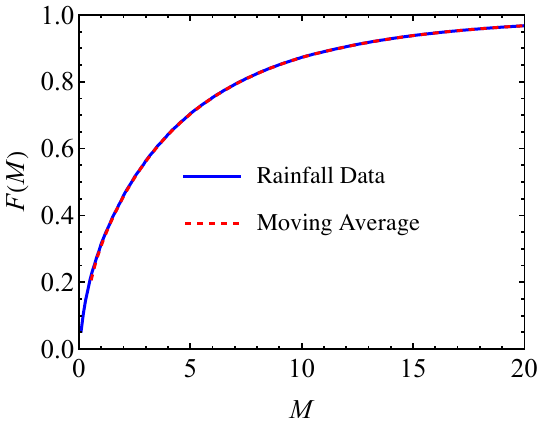}
    \caption{(Left) Empirical probability distribution function (PDF) $f(x)$ obtained from the combined rainfall data of the wet days for the period 1974--2013, restricted to the data for which $q$ lies in
the interval $[0.48, 0.52]$. For the combined dataset, the
fraction of dry days is $q \approx  0.5016 > q_c \approx 0.3152$. The blue points are computed directly from the data, and the red points show moving average taken over 20 bins. (Right) The cumulative distribution function (CDF) $F(M) = \int_0^M f(x)\, dx$ constructed from the same data. The blue line is obtained from the original data and the red dashed line shows moving average over 10 points. The CDF from the original data is indistinguishable from its moving average. The CDF $F(M)$ is the one (rather than its PDF $f(x)$) that appears in the calculation of $C_q(n)$. }
    \label{fig:pdf_cdf}
\end{figure}

We estimate the empirical PDF $f(x)$ of daily precipitation amounts $x$
(in mm) on wet days from a normalized histogram with a bin width of
$0.1$\,mm, equal to the resolution of the DWD records. Because the resulting PDF is noisy, we
smooth it with a moving average over 20 consecutive bins. Both the raw PDF
and its smoothed version are shown in Fig.~\ref{fig:pdf_cdf} (left).

To compute the covariance $C_q(n)$ defined in Eq.~\eqref{def_Cq}, we need
the cumulative distribution function (CDF) $F(M)$ of wet-day amounts,
defined in Eq.~\eqref{A1.1}. As Fig.~\ref{fig:pdf_cdf} (right) shows, the
empirical CDF is much smoother than the PDF, and the raw CDF and its moving
average are nearly indistinguishable. Numerical integration of the empirical
CDF gives $\mu_1 = \int_0^\infty [1 - F(M)]\, dM \approx 4.62314$ and
$\mu_2 = \int_0^\infty [1 - F(M)]^2\, dM \approx 1.83823$. Therefore, using the expression of $q_c$ (see main text and End Matter)
\bea \label{cond_qc}
q_c =\frac{1}{2} \left(- \left(1+\frac{2\mu_1}{\mu_2}\right) + \sqrt{\left(1+\frac{2\mu_1}{\mu_2}\right)^2+8}\right) \;,
\eea
we get $q_c \approx 0.3152$. Since $q > q_c$ for
our data, we expect $C_q(n)$ to be non-monotonic in $n$.

For each $n$, we compute $C_q(n)$ by extracting the coefficient of $v^n$ in
the integrand of Eq.~\eqref{GF_Cq_final} and integrating it numerically over
$M \in (0,\infty)$, with $g(M) = q + (1-q)F(M)$ taken from the data. As shown
in Fig.~3 (left) of the main text, this theoretical $C_q(n)$ agrees very well
with $C_q(n)$ measured directly from the observational data when 12 days are
skipped between consecutive observations, which renders them
effectively uncorrelated.

\begin{figure}
    \centering
    \includegraphics[width=0.48\textwidth]{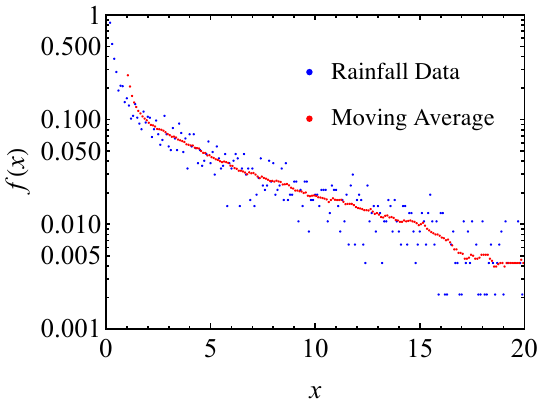}~~\includegraphics[width=0.46\textwidth]{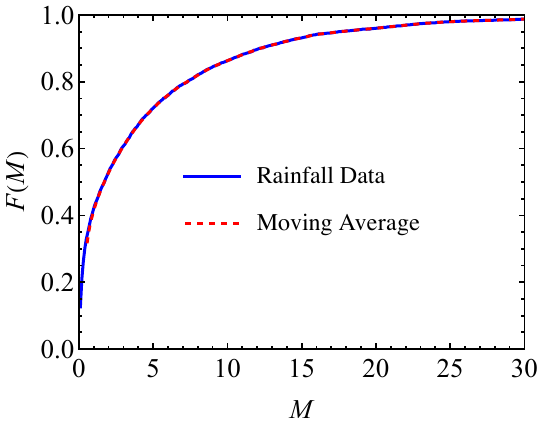}
    \caption{(Left) Empirical probability distribution function (PDF) $f(x)$ obtained from the combined rainfall data of the wet days for November from the data for which the dry-day fraction does not exceed 0.275. The effective dry-day fraction for this subset turns out to be $q \approx 0.2323 < q_c \approx 0.2795$. The blue points are computed directly from the data, and the red points show moving average taken over 20 points. (Right) The cumulative distribution function (CDF) $F(M) = \int_0^M f(x)\, dx$ constructed from the same data. The blue line is obtained from the original data and the red dashed line shows moving average over 10 points. The CDF from the original data is indistinguishable from its moving average. The CDF $F(M)$ is the one (rather than its PDF $f(x)$) that appears in the calculation of $C_q(n)$. }
    \label{fig:pdf_cdf2}
\end{figure}

To examine small values of $q$, we restrict the analysis to November and
select 15 stations from the full dataset that have complete November records
and a November dry-day fraction of at most $0.275$. The dry-day fraction for
the combined data from these stations is $q = 0.232353\ldots$. The empirical
PDF and CDF for this dataset are shown in Fig.~\ref{fig:pdf_cdf2}. The raw PDF
is noisier than in the previous case, but the CDF remains reasonably smooth.
Following the same steps as above, we find $\mu_1 \approx 4.40274$ and
$\mu_2 \approx 1.49851$. Using $\mu_1$ and $\mu_2$ in Eq.~\eqref{cond_qc}, we get $q_c \approx 0.2795$. Since $q < q_c$, we expect
$C_q(n)$ to decay monotonically. Figure~3 (right) of the main text compares
this theoretical $C_q(n)$ with $C_q(n)$ measured directly from the
observational data, now with 10 days skipped between consecutive
observations.

\newpage

\appendix